\documentclass[fleqn,usenatbib]{mnras}

\usepackage{newtxtext,newtxmath}

\usepackage[T1]{fontenc}

\DeclareRobustCommand{\VAN}[3]{#2}
\let\VANthebibliography\thebibliography
\def\thebibliography{\DeclareRobustCommand{\VAN}[3]{##3}\VANthebibliography}

\usepackage{graphicx}	
\usepackage{amsmath}	
\usepackage{float} 
\usepackage{soul,color}
\soulregister\cite7
\soulregister\ref7
\soulregister\pageref7

\title[M82 starburst and outflow]{Revisiting the Galactic Winds in M82 I: the recent starburst and launch of outflow in simulations}

\author[T. R. Wang et al.]{
Tian-Rui Wang,$^{1}$
Weishan Zhu,$^{1}$\thanks{E-mail: zhuwshan5@mail.sysu.edu.cn (WSZ)}
Xue-Fu Li, $^{1}$ 
Wen-Sheng Hong$^{2}$
and Long-Long Feng$^{1}$
\\
$^{1}$Department of Astronomy, Sun Yat-Sen University, No. 2 Daxue Road, Xiangzhou District, Zhuhai, 519082, China\\
$^{2}$Department of Astronomy, School of Physics and Astronomy, and Shanghai Key Laboratory for Particle Physics and Cosmology, \\~~~~Shanghai Jiao Tong University, Shanghai 200240, People's Republic of China\\
}

\date{Accepted XXX. Received YYY; in original form ZZZ}

\pubyear{2015}

\begin{document}
\label{firstpage}
\pagerange{\pageref{firstpage}--\pageref{lastpage}}
\maketitle

\begin{abstract}

We revisit the launch of the galactic outflow in M82 using hydrodynamic simulations. Employing a sink-particle module, we self-consistently resolve star formation and feedback, avoiding reliance on simplified models. We investigate the effects of stellar feedback mechanisms, gas return from star-forming clouds, and disk mass on the starburst and outflow. Our simulations generate a starburst lasting $\sim25$ Myr, peaking at 20–50 $\rm{M_{\odot},yr^{-1}}$, although the total stellar mass often exceeds M82's estimated value. The outflow develops in two stages: initially, continuous SNe form small bubbles that merge into a superbubble containing warm/hot gas and intermediate- to high-density cool filaments. After $\sim10$ Myr, the superbubble breaks out of the disk, and within $\sim15$ Myr a kpc-scale outflow forms. Cool filaments survive stellar feedback, become entrained in the wind, and stretch to hundreds of parsecs. Transport from the cool ISM is the dominant net contributor to the total mass of the cool phase in the outflow, whereas transfers from hotter phases, such as through condensation or precipitation, provide only a minor net contribution, likely offset by simultaneous transfer from the cool phase back to hotter phases. While the mass loading factor is comparable to M82, the cool gas outflow rate and velocity are lower, with velocities $\sim60\%$ below observed values; warm and hot gas are $\sim25\%$ slower. SN feedback is the primary driver, and gas return significantly influences the starburst and outflow, while other factors are secondary. Stronger clustered SN feedback is likely required to better match observations.

\end{abstract}




\section{Introduction}

Since the initial discovery in M82 by \cite{1963ApJ...137.1005L}, galactic winds and outflows have been observed in both nearby (e.g., \citealt{1987AJ.....93..264M,1990ApJS...74..833H,1999ApJ...513..156M,2000ApJS..129..493H,2005ApJS..160..115R,2014A&A...568A..14A,2015ApJ...809..147H,2019MNRAS.483.4586F,2023arXiv231006614X}) and high redshift universe(e.g. \citealt{2009ApJ...692..187W,2012ApJ...760..127M,2014ApJ...796....7G,2023ApJ...959..124K,2023ApJ...949....9P,2024A&A...685A..99C,2024MNRAS.531.4560W}). Observations spanning multiple bands have revealed that galactic winds are complex, multi-phase structures that extend from a few parsecs to tens of kiloparsecs. These outflows contain a mixture of gas phases, including cold ($\rm{T}\lesssim 10^3\,$K), cool ($\rm{T}\sim 10^3-2\times10^4\,$K), warm ($2\times10^4\,\rm{K} \leq \rm{T} < 5\times10^5$K), hot ($5\times10^5\leq\rm{T}<10^7\,$K), and very hot ($\rm{T}\geq10^7\,$K) gaseous, and dust (for more details, see reviews by \citealt{2005ARA&A..43..769V, 2020A&ARv..28....2V, 2017arXiv170109062H, 2018Galax...6..138R} and references therein). Observations, analytical models, and simulations have established that stellar feedback, including supernovae, stellar winds, radiation, and possibly cosmic rays, and active galactic nuclei (AGN) feedback can drive galactic-scale outflows, reproducing observed multi-phase structures  (e.g., \citealt{1985Natur.317...44C,Strickland2000,2005ApJ...618..569M}). Extensive research suggests that these outflows play a crucial role in the redistribution of the interstellar medium (ISM), regulating the efficiency of star formation and enriching the circumgalactic medium (CGM) and intergalactic medium (IGM) (\citealt{1986ApJ...303...39D,1994MNRAS.271..781C,2017ARA&A..55...59N}). This work focuses on stellar feedback driven outflows, widely considered to be the primary mechanism for suppressing star formation in low-mass galaxies. 

Despite significant progress over the past six decades, many fundamental questions about starburst driven galactic winds remain unanswered. Observationally, precisely measuring the key properties of outflows, including velocities, mass, momentum, and energy outflow rates, particularly for warm, cool, and cold phases, remains challenging (\citealt{2017arXiv170109062H, 2018Galax...6..138R, 2020A&ARv..28....2V}). Theoretically, while analytical models and simulations can explain hot winds, the origin of cool gas and the mechanisms accelerating cool gas to hundreds of km/s remain uncertain (e.g. \citealt{2018MNRAS.480L.111G,2022ApJ...924...82F}). The roles of radiation pressure and cosmic rays in boosting galactic winds and accelerating warm and cool phases are also open questions (\citealt{2017arXiv170109062H, 2018Galax...6..114Z}). Furthermore, simulations of  galactic winds in galaxies with similar stellar and gas mass can yield divergent results due to variations in simulation setups, including SN feedback prescriptions, resolution, and initial ISM conditions.

A deeper understanding of how stellar feedback initiates galactic winds and the underlying processes driving their evolution is essential. Such knowledge will not only help interpret growing observational data but also solidify the modern $\Lambda$CDM paradigm of galaxy formation and evolution (e.g.,\citealt{2017ARA&A..55...59N}). Within this framework, galactic winds are crucial for suppressing star formation in galaxies below the characteristic luminosity $\rm{L_*}$, preventing overproduction of faint galaxies (e.g. \citealt{1986ApJ...303...39D,1994MNRAS.271..781C, 2015ARA&A..53...51S}). Recent cosmological simulations of galaxy formation and evolution, such as Horizon-AGN (\citealt{2014MNRAS.444.1453D}), EAGLE (\citealt{2015MNRAS.446..521S}), IllustrisTNG (\citealt{2018MNRAS.473.4077P}), Simba (\citealt{2019MNRAS.486.2827D}), have incorporated galactic winds and AGN feedback to successfully reproduce many observed galaxy properties. However, despite addressing similar physical processes, these simulations employ diverse sub-grid models for stellar feedback and galactic winds. EAGLE utilizes stochastic thermal heating from massive stars, Horizon-AGN adopts mechanical feedback (\citealt{2017MNRAS.467.4739K}), and IllustrisTNG employs isotropic kinetic winds. Furthermore, the values of key parameters, including kinetic energy fraction, velocity, and mass outflow rate, vary significantly across these codes. 

To gain a deeper understanding of galactic winds, a systematic theoretical investigation across multiple scales is necessary. On a parsec scale, detailed studies are required to explore the formation and lifetime of cold gas clumps and filaments in hot winds (e.g. \citealt{2009ApJ...703..330C, 2015MNRAS.449....2M, 2015ApJ...805..158S, 2018MNRAS.480L.111G}), thermal instability in multi-phase winds, dust-gas coupling, and the efficiency of radiation pressure and cosmic rays, using both analytical models and simulations.
Concurrently, global models and simulations spanning kpc scales are needed to examine the combined effects of these processes in realistic galaxy environments. Nearby galaxies with extensive observations, such as M82, provide ideal laboratories for testing these global simulations.

Numerous simulations of M82-like galaxies have been conducted over the past four decades. Early studies often employed one- and two-dimensional, axisymmetric simulations to investigate the development of galactic winds driven by nuclear starbursts in isothermal interstellar medium (e.g. \citealt{1988ApJ...330..695T, 1994ApJ...430..511S,1996ApJ...463..528S, 1998MNRAS.293..299T, 1999ApJ...513..142M, Strickland2000, 2003ApJ...597..279T, 2009ApJ...697.2030S}). These idealized simulations have validated the \cite{1985Natur.317...44C} model in the adiabatic phase and demonstrated that the properties of the wind, such as dynamics, morphology, and X-ray emission, are influenced by factors such as the distribution of the interstellar medium, the ambient gaseous halo, and the distribution and clustering of young stars. 

More recently, three-dimensional simulations (e.g. \citealt{2008ApJ...674..157C,2008ApJ...689..153R,  2013MNRAS.430.3235M}) have been employed to model the galactic wind in M82-like galaxies. \cite{2008ApJ...674..157C} suggested that H-$\alpha$ filaments form from disk clouds entrained and stretched by the hot wind, emphasizing the role of the inhomogeneous disk in shaping the wind's filamentary structure. \cite{2008ApJ...689..153R} and \cite{2013MNRAS.430.3235M} highlighted the importance of radiative interactions between winds from super star clusters (SSCs), as well as the number, distribution, and buildup of SSCs, in developing H-$\alpha$ filaments. Very recently, \cite{2020ApJ...895...43S} and \cite{2024ApJ...966...37S} conducted three-dimensional simulations of outflows driven by clustered supernova feedback in a large domain ($\sim20$ kpc) with high resolution ($\sim5$ pc). These simulations successfully produced multiphase outflows that closely resemble M82's outflow in many aspects. 

However, as most prior simulations of galactic winds in M82-like galaxies focus on how different physical mechanisms influence the overall properties of outflows, they often adopt some simplified assumptions and models, which limit their ability to compare directly to M82. A common factor is that the recent starburst in the central disk region is not resolved self-consistently. Instead, previous simulations often manually introduce star clusters with varying total masses (1 $\times 10^{8}$ to 6 $\times 10^{8} \rm{M_\odot}$), spatial distributions, and age distributions in different studies (\citealt{2008ApJ...674..157C,2008ApJ...689..153R,  2013MNRAS.430.3235M,2020ApJ...895...43S}).  Additionally, assumptions about the M82-galaxy properties, such as its mass model and gas distribution, are often outdated from recent observations of M82, which have revealed discrepancies with earlier studies, particularly regarding the rotation curve ( \citealt{2012ApJ...757...24G}), dark matter halo and gas distribution (e.g., \citealt{Adam2015}). Furthermore, the resolution and numerical methods employed in many previous simulations, including stellar evolution, radiative cooling, radiative pressure, and heating, can be further refined.

To enhance our understanding of galactic winds, we revisit M82's outflows using high-resolution three-dimensional hydrodynamic simulations that incorporate updated knowledge of M82, a more realistic representation of the recent nuclear starburst, and refined numerical treatments of relevant physical processes. This work, the first in a series, focuses on resolving the nuclear starburst and the subsequent launch of kpc-scale galactic outflows. Our second work (\citealt{2025ApJ...982...28L}) delves into the development of multiphase outflows and the impact of various factors on their properties. This paper is organized as follows. Section 2 outlines our methodology. Section 3 presents the properties of the starburst in our simulations. The launch of kpc-scale multiphase outflows is described in Section 4. Section 5 discusses our findings, including comparisons with previous work and limitations. Finally, Section 6 summarizes our findings.

\section{Method}
\label{sec:Method}
Section 2.1 outlines our adopted mass model for M82. Sections 2.2 and 2.3 detail the initial conditions and star formation modules, respectively. Section 2.4 describes the implementation of various feedback mechanisms. Radiative cooling and heating processes are discussed in Section 2.5. Finally, Section 2.6 provides an overview of the simulations presented in this work.

\subsection{Mass model of M82}
Recent observational advances have provided more detailed estimates of the mass and distribution of baryonic matter, including stars and various gas phases. Furthermore, recent observations of the rotation curve suggest the presence of a dark matter halo, which was not considered in earlier studies. We construct a mass model for M82 that incorporates these latest observations. While significant uncertainties remain in the masses of gas and stars, we account for these by varying parameters in our models. Our mass models generally follow the approach of previous studies (e.g. \citealt{Strickland2000,2008ApJ...674..157C,2013MNRAS.430.3235M,2020ApJ...895...43S}). Our mass models mainly account for the contribution of a stellar disk, a gas disk, and a dark matter halo. The latter is not considered in earlier works such as \cite{Strickland2000,2008ApJ...674..157C,2013MNRAS.430.3235M}, but is needed to reproduce the latest observed rotation curve of M82 in \cite{2012ApJ...757...24G}. 

We adopt a \citealt{Miyamoto1974} profile for the stellar disk, with a mass of $ 3.3\times 10^9\, \rm{M_{\odot}}$, a scale length of 1200 pc, and a scale height of 200 pc. M82 also consists of a bulge with a mass of $\sim 5.0 \times 10^8 \rm{M_{\odot}}$ in the central 1 kpc region. A considerable fraction of stars in the bulge is likely formed in the recent star bursts, considering the spatial overlapping, and since this work focuses on simulating the recent starburst, we incorporate the bulge mass into the gas component, that is, the gas disk 2 in Table \ref{tab:disk&halo}. The gas disk is modeled using a triple Miyamoto-Nagai profile similar to (\citealt{Smith2015}), with a mass of $\rm{M_{gd}} \approx 2.5, 3.0, 3.5 \times 10^9 \rm{M_\odot}$, and a half-mass radius of 1 kpc and a ellipticity of 0.3. The dark matter halo in our model follows a core-NFW profile with a mass of $\rm{M_{dmh}} \approx 6.0\times 10^{10}\, M_{\odot}$. We manage to determine the related parameters for each component so that the total rotation curve predicted by these components can reproduce the result in \cite{2012ApJ...757...24G}, as Figure \ref{fig:rotcurve} shows. Detailed information about the mass model is provided in Appendix A. 

\begin{figure}
    \centering
    \includegraphics[width=\columnwidth]{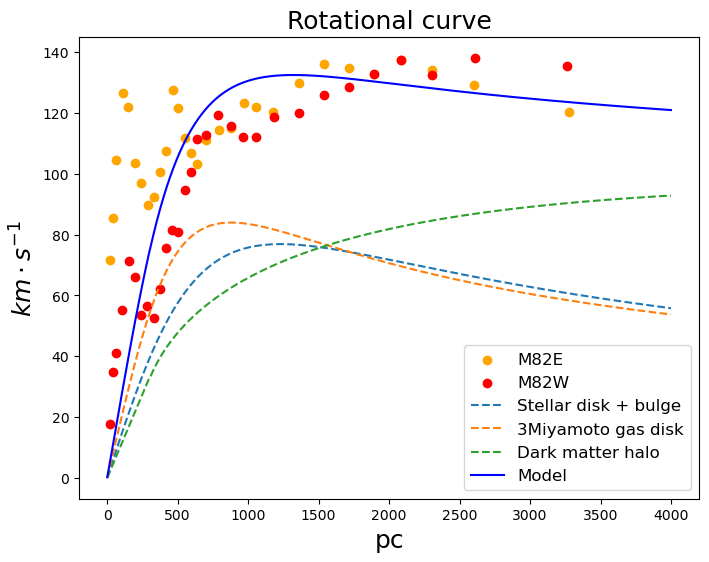}
    \caption{Observation results of the rotation curve provided by \citealt{2012ApJ...757...24G}(red and yellow dots) and our model (blue solid line). The rotation curve contributed by the different components is shown as dashed lines.}
    \label{fig:rotcurve}
\end{figure}

\subsection{Initial conditions in simulation}
In our simulation, we first set up the gas distribution in each grid cell at the beginning of the simulation and then track the evolution of gas density and temperature, as well as star formation and feedback processes. The gravitational effect of the stellar disk and the dark matter halo is incorporated as a background field, a reasonable assumption for our 30 Myr simulation duration. The self-gravity of gas and newly formed stars is solved at each time step by the module in Athena++ (\citealt{Stone2020}).  

The density profile of the gas disk component in the state of hydrostatic equilibrium under the galaxy gravity potential is given by 
\begin{equation}
\begin{split}
    \rho(r,z) = & \rho_{0} \  \times \\
    & \rm{exp}\left[ -\frac{\Phi_{tot}(r,z)-e^{2}\Phi_{tot}(r,0)-(1-e^{2})\Phi_{tot}(0,0)}{c_{\rm{{s,eff}}}^{2}} \right], 
\end{split}
\label{eqn:gas_disc_prof}
\end{equation}
where $\rho_{0}$ is the central gas density, the rotational factor e reads as
\begin{equation}
    e = e_{\rm{rot}}\,\rm{exp}(-z/z_{rot})
\end{equation}
where $e_{\rm{rot}}$ in the above equation is chosen to be 0.92 for a thick disk. For the scale height $\rm{z}_{rot}$, we choose a value of $\rm{z}_{rot} = 5\, \rm{kpc}$, following the literature (\citealt{1994ApJ...430..511S, Strickland2000,2008ApJ...674..157C}). In addition to the gas disc, we also include a hot gas halo that follows a profile similar to that of Eqn. \ref{eqn:gas_disc_prof} but with a much smaller $e_{\rm{rot}}$ , which is consistent with previous studies(\citealt{Strickland2000, 2008ApJ...674..157C}). The parameters for the isothermal disk and the hot gas halo are listed in Table \ref{tab:disk&halo}. Note that $c_{\rm{s,eff}}$ stands for the effective sound speed that correspond to the combined energy density of both the non-thermal pressure provided by turbulence and the static pressure provided by internal energy.

\begin{table}
	\centering
	\caption{The initial central density, $\rho_{0}$, rotation factor, $e_{\rm{rot}}$, and effective sound speed of the gas disc and halo. Gas disc 1 represents the dominate gas disc component. Gas disc 2 actually is used to mimic the potential of bulge, see text for more details.}
	\begin{tabular}{lcccr}
		\hline
		Component & $\rho_{0}$ & $e_{\rm{rot}}$  & $c_{\rm{s,eff}}$\\
		\  & $cm^{-3}$  & &$cm\cdot s^{-1}$\\
		\hline
		gas disk 1 (extended disk) & 190 & 0.92 & $3.6\times 10^{6}$\\
        gas disk 2 (bulge) & 70 & 0.5 & $5.7 \times 10^{6}$\\
		hot gas halo & 2.0e-3 & 0.25 & $3.0 \times 10^{7}$\\
		\hline
	\end{tabular}
	\label{tab:disk&halo} 
\end{table}

Observations reveal a turbulent interstellar medium (ISM) in M82, with non-thermal velocity dispersions of $\sigma \sim 20 \, \rm{km\ s^{-1}}$ in molecular clouds (\citealt{Westmoquette2013}). Turbulence likely plays a significant role in both star formation and disk support. To account for this, we introduce turbulence into the initial gas disk. Unlike \cite{2008ApJ...674..157C}, who employed a static, inhomogeneous ISM with a lognormal probability density function, we perturb the velocity field using an energy spectrum function, given by
\begin{equation}
    E_{k}(k) \propto k^{-p}, 
\end{equation}
and
\begin{equation}
     \left< (\delta u(r))^{2} \right>  \propto r^{p-1}.
\end{equation}
For typical Kolmogorov turbulence, $p = \frac{5}{3}$. However, turbulence in the ISM is generally supersonic (e.g.\citealt{2004ARA&A..42..211E,2010A&A...512A..81F}), with $p$ closer to 2 (e.g.\citealt{2002ApJ...569..841B,2013MNRAS.436.1245F}). Given the mean velocity dispersion reported by \cite{Westmoquette2013}, the ISM in the star-forming region of M82 is in a supersonic turbulent state. We initialize the velocity of each grid cell as the superposition of rotational and turbulent components, and the latter is assumed to have a velocity energy spectrum with $p=2$, with a velocity dispersion magnitude $\sigma=20 \, \rm{km\ s^{-1}}$. The velocity field of the turbulence adopted in our simulation is calculated using the Fourier transform. It is then applied only to the gas disk component. 

Then, to set up a stable disk, the static pressure of the gas disc is determined by subtracting the non-thermal pressure provided by turbulence from the effective pressure that calculated using the effective sound speed. The non-thermal pressure provided by turbulence takes the form of
\begin{equation}
    p_{nt}(v_{tur}) = \frac{1}{2} \rho {v_{tur}}^{2}
\end{equation}
while the pressure of ideal gas takes the form of
\begin{equation}
    p(c_{s}) = \frac{1}{\gamma} \rho c_{s}^{2}
\end{equation}
The static pressure of gas in disc is then given by
\begin{equation}
    p_{\rm{static}} = p(c_{\rm{s\_eff}}) - p_{nt}(v_{tur})
\end{equation}

For gas in the hot halo, turbulence is absent, and the static pressure is determined by the effective sound speed. At each grid point, the total gas density is given by the sum of the densities of the three components listed in Table \ref{tab:disk&halo}, while the static pressure is obtained by summing the static pressures of these components. The gas distribution is then evolved without cooling or feedback for 2.5 Myr to generate an inhomogeneous disk, which serves as the initial condition for our simulations. 

\subsection{Star formation}
Previous simulations of galactic winds in M82-like galaxies often rely on prescribed star formation histories, such as instantaneous starbursts or constant star formation rates. Additionally, SN feedback has been injected into a thin disk plane (\citealt{Strickland2000, 2008ApJ...674..157C}), or a central 1 kpc spherical region based on assumed star cluster distributions (e.g.,\citealt{2013MNRAS.430.3235M};\citealt{2020ApJ...895...43S};\citealt{2024ApJ...966...37S}). A key goal of our work is to more self-consistently model the recent starburst in M82's nuclear region and its associated feedback. To this end, both the star formation process and how they couple with the surrounding ISM are solved explicitly in our simulation. 

Newly born stars are embedded in molecular clouds and are usually surrounded by overdense ISM. Recent simulations of dwarf galaxies, which resolve star formation down to the scale of molecular clouds, have employed various approaches to capture star formation process. For dwarf galaxies with masses of $10^7-10^8 \rm{M_\odot}$ and relatively low SFR, a star-by-star formation recipe (\citealt{Thales2021}), is effective when sub-parsec resolution is achieved. However, for galaxy-scale simulations with coarser resolution (several parsecs or larger), a sink particle approach is more practical. This approach involves creating sink particles with masses exceeding that of a single star in high-density regions, allowing them to accrete mass from surrounding gas cells until they meet the criteria for spawning star particles, each representing a cluster of stars (e.g., \citealt{1995MNRAS.277..362B,2004ApJ...611..399K,2005A&A...435..611J, Federrath2010, Gong2012, Gatto2016, Howard2016}). We adopt a sink particle model based on the methods of \cite{Federrath2010} and \cite{Howard2016} to solve the star formation process in our work..

We have implemented a sink particle solver within the Athena++ code (\citealt{Stone2020}) to handle star formation and evolution. At each time step, gas cells are inspected to determine if they meet the following criteria for forming a sink particle:
\begin{enumerate}
    \item Jean's instability: The gas cell must be gravitationally unstable.
    \item does not overlap with existing sink/star particle.
    \item convergent gas flow.
    \item located at the center of a dense gas clump.
\end{enumerate}

These criteria are commonly used for sink particle implementations in mesh-based fluid dynamic solvers (e.g. \citealt{2004ApJ...611..399K, Federrath2010, Gong2012}). The criterion requiring the cell to be gravitationally bound has also been adopted in many studies. However, some studies indicate that stars can form in areas that are not gravitationally bound, such as the fast-expanding shell of an SNR bubble or a wind bubble (\citealt{Tristan2018, Deharveng2009}) or high-velocity fragments in the outflow area (\citealt{Maiolino2017}). To account for these scenarios, we relax the gravitational bound criterion in our simulations. The following subsections detail these criteria and the star formation process from sink particles.

\subsubsection{Jean's instability}
A fundamental requirement for star formation is Jeans instability. In a mesh-based hydrodynamical code, this translates to a density threshold, as proposed by \cite{Truelove1997}:
\begin{equation}
    \rho_{th} = \frac{\pi}{16} \frac{c_{s}^{2}}{G\Delta x^{2}},
\end{equation}
where $c_{s}$ is the local sound speed and $\Delta x$ is the cell size. However, as noted by \cite{Federrath2010}, Truelove's criterion is only applicable to regions undergoing free-fall collapse. To identify such regions, further analysis is necessary, as detailed in the following subsections.

\subsubsection{Distance check}
Even if gas clouds near a newly formed star cluster meet the density threshold for star formation, they can be disrupted by photoionization heating and radiative pressure from massive stars. In M82, star clusters are often embedded in molecular clouds with optical depths of up to $\tau \sim 10$. In such environments, radiative pressure may be comparable to supernovae in terms of momentum injection (\citealt{Agertz2012}). Combined with stellar winds, these early feedback processes can suppress star formation in neighboring regions. To account for this, we assign each sink particle a control volume that should not overlap with those of existing sink or star particles. The diameter of these spherical control volumes is set to three times the grid cell size.
 
\subsubsection{Convergent flow}
To confirm that a candidate cell is undergoing free-fall collapse, we check for convergent gas flow. This requires that the divergence of the velocity field , $\nabla \cdot v < 0$, is negative in the central cell of the control volume (\citealt{Gong2012}).

\subsubsection{Gravitational potential minimum}
A sink particle is created in a candidate cell only if it corresponds to a local minimum of the gravitational potential (\citealt{Federrath2010}), $\Phi$. In practice, we consider only the gravitational potential of the gas, $\Phi_{gas}$, when examining this criterion. Neglecting the background potential from the stellar disk and dark matter halo prevents the sink particle from being placed away from the densest region of the gas clump.

\subsubsection{Gas accretion and star formation}
Several methods exist for calculating gas accretion onto sink particles. \cite{2004ApJ...611..399K} categorize accretion into three scenarios based on the relative importance of pressure and gravity, as determined by the particle's Bondi-Hoyle radius, and employ different accretion models accordingly. \cite{Gong2012}  introduce `ghost' zones around particles and calculate accretion rates based on the flux at the zone boundaries. In our work, we adopt a simpler approach similar to \cite{Federrath2010}. Once a sink particle forms, gas accretion begins immediately. Grid cells within the control volume are continuously monitored, and any gas exceeding the density threshold $\rho_{th}$ is directly accreted by the sink particle. The mass accretion rate of a sink particle at each time step is given by: 
\begin{equation}
    \Delta M = \sum (min(\rho(i,j,k) - \rho_{th}(i,j,k)), 0) \Delta V(i,j,k), 
\end{equation}
where the summation is taken over all the gas cells within the control volume of this sink particle. Such an implementation of gas accretion enforces Turelove's stability criterion, as well as the conservation of mass. Furthermore, this scheme is reasonable because the Bondi-Hoyle radius (\citealt{Bondi1952}), 
\begin{equation}
    r_{BH} = \frac{GM}{v^{2}+c_s^{2}}
\end{equation}
for a molecular cloud with a mass around $10^{4}\sim 10^{5} \rm{M_{\odot}}$ and a sound speed of $\sim 1 \rm{km\cdot s^{-1}}$ and is larger than the resolution of the most refined grids in our simulation. The inflow toward a sink particle is therefore supersonic, making the specific choice of accretion rate proposed by \cite{2004ApJ...611..399K} irrelevant. In addition to mass conservation, momentum and metal mass are also conserved during accretion. The sink particle's velocity is updated at each timestep using momentum conservation, and the accreted gas's metallicity is assigned as the initial metallicity for newly formed stars.

We employ a widely used method to convert the gas accreted to sink particles into stellar mass at a rate of 20\% per free-fall time. This conversion rate ensures that the surface star formation rate scales with density as $\rho^{1.5}$, consistent with the Kennicutt-Schmidt law (\citealt{Thales2021}). The star formation rate for a sink particle at time t is given by:
\begin{equation}
    \label{eq:mass_convert}
    \dot M_{*} (t) = \frac{M_{gas}(t)}{t_{SF}},
\end{equation}
where $M_{gas}(t)$ is the gas mass stored in this sink particle and 
\begin{equation}
    t_{SF} = \frac{1}{0.2} t_{ff} = \frac{1}{0.2} \sqrt{\frac{3 \pi}{32 G \rho }} . 
\end{equation}
During our simulation, the star formation time scale $t_{SF}$ will be updated after accretion at each time step. Based on newly updated $t_{SF}$, a fraction of the gas stored in the reservoir will be converted to stellar mass following Equation \ref{eq:mass_convert} within each time step, and then the stellar mass of each sink particle is updated. In this way, a sink particle represents both a gravitationally bound star-forming gas clump and the embedded stars within it (\citealt{Howard2016}). 

Initially, the stellar mass within a star-forming gas clump grows gradually. However, once the stellar mass in a gas clump exceeds some critical value, the gas accretion would cease due to feedback from newly born stars. In our simulation, if the stellar mass in a sink particle exceeds a threshold value of $M_{cnv}$, it will stop accreting gas and be reclassified as a star particle. The mass of star clusters in M82 can vary from $1000\,\rm{M_{\odot}}$ to as high as $10^{6}\, \rm{M_{\odot}}$. Observations show that the median mass of super star clusters (SSCs) in M82 is around $10^{4.12}\, \rm{M_{\odot}}$ and the median radius is around 4.26 pc (\cite{Bolivia2020}). 

In our simulation, the radius of the control volume of a sink particle, $R_{cv}$, is $ 1.5$ times the size of the nearby grid cell. Since sink particles can form in regions with varying resolutions, their control volumes and the corresponding Jeans masses $M_{J}$ can differ. A fixed mass threshold, $M_{cnv}$, could lead to inconsistent distribution of stellar mass fraction, i.e. the ratio of stellar mass to the total mass in a star particle, across regions with different resolutions. To ensure a consistent distribution of the stellar mass fraction, we adopt a variable $M_{cnv}$. Based on the assumption that a) the total accreted mass is proportional to Jeans mass, b) sink particles of different sizes have a constant temperature, and c) the stellar mass fraction is independent of the particle mass, we derive $M_{cnv} \propto M_{J} \propto R_{cv}$. In our simulation, we adopt $M_{cnv} = 5.0\times 10^{4} \,\rm{M_{\odot}}$ in regions have the finest grids with size $dx = 4.0$ pc, and $M_{cnv} = 1.0\times 10^{5} \,\rm{M_{\odot}}$ in regions with $dx = 8.0$ pc, etc. 

Under this star-formation scheme, the accreted gas is not fully converted into stars. Once the stellar-mass threshold is reached, the sink particle is converted into a star particle, which represents a stellar cluster containing both the newly formed stars and the remaining gravitationally bound gas. The detailed treatment of this residual gas is described in Section \ref{sec:gr}.

\subsection{Feedback}
Stellar feedback serves as the main driving force for the launch of the galactic outflow. Traditionally, the primary feedback mechanism considered in similar studies is the core-collapse SNe, which has been proven capable of launching galaxy-scale outflow in M82-like galaxies (e.g. \citealt{Strickland2000, 2008ApJ...674..157C, 2020ApJ...895...43S}). However, recent studies show that radiation feedback and stellar wind may play crucial roles in manipulating the effect of SNe feedback (\citealt{Dale2014,2020MNRAS.493.2872C}). These early feedback processes may significantly lower the SFR and prevent the clustering of star formation. Some previous work suggests that efficient X-ray and extreme ultraviolet (EUV) radiation are capable of removing important coolants through photoionization, leading directly to the reduction of the accretion rate and SFR (e.g. \citealt{Sebastiano2009}). Furthermore, stellar wind can create cavities and channels of ionized gas before the first SNe, mitigating the resistance an expanding SNR may meet. In this study, we consider three major feedback processes: radiation, stellar wind, and core-collapse SNe. We will examine how these mechanisms affect the starburst activity and the launch of outflow in M82. The effects of Type Ia SNe are ignored, as our simulations span only 30 Myr, during which their contribution to the feedback energy is expected to be minor compared to that of core-collapse supernovae. In our simulations, the stellar mass of a single star particle is equivalent to a medium-sized star cluster. Therefore, it can continuously produce stellar wind and radiation feedback and it will launch several individual SNe throughout its lifetime. 
 
On the other hand, not all the gas in sink/star particles (GMCs) will convert to stars. Stellar feedback can disperse a great amount of the remaining gas in GMCs, causing a considerable amount of gas to return to the diffuse ISM. Studies indicate that up to $80\%$ of the mass of star-forming gas clumps will return back to the ambient diffuse ISM within the time scale of $5 \sim 15 $ Myr, equivalent to several to ten times of the newly born stellar mass,(e.g., \citealt{Dale2014, Kim2018, li2019disruption, Fujii2021}). This process of gas return can have substantial effects on the initiation and development of outflows. The detailed implementations of gas return and the three stellar feedback processes mentioned above are also introduced in this subsection, and their effects will be presented in later sections.

\subsubsection{Gas return from sink particle}
\label{sec:gr}
 Most previous studies that adopted a sink particle-based star formation scheme have not included any returning mechanism for the remaining gas in sink particles. Some of these studies may assume that all the gas in sink particles will fully convert to stars (e.g. \citealt{2010ApJ...709...27W,2011ApJ...730...40P, Gong2012, Kim2018}), which is valid only if the resolution can reach 0.1 parsec or higher. However, this treatment may overestimate the star formation efficiency (SFE and hereafter, i.e., the fraction of gas in GMCs/galaxy that has been converted into stars within a given time interval) in simulations with a resolution coarser than the size of the cores of GMCs. Meanwhile, other simulation studies on the evolution of GMC and ultra-faint dwarf galaxies believe that most unresolved gas in sink particles remains gravitationally bound to the cluster formed in GMC over a long time scale (\citealt{Dale2014, Howard2016}). Therefore, gas is stored in sink particles in these studies. However, in simulations with a relatively coarser resolution or with a long simulation time, such a treatment would cause a great fraction of the gas to be locked up in particles (\citealt{Truelove1997, Hu2017}), leading to an underestimation of the star formation efficiency. Overall, both approaches mentioned above have their own limitations. Recent studies indicate that early pre-supernovae feedback processes (i.e. stellar wind, radiation feedback) are capable of destroying the cloud and accelerating a great portion of the remaining gas to a gravitationally unbound state within 2 $\sim$ 15 Myr, in which the returning fraction may depend on the properties of the gas clumps (e.g., \citealt{Kim2018,li2019disruption}). 
 
 In our simulation, we use a simplified model to account for the dispersal of gas remaining in the sink/star particles. In the default run, we assume that all the remaining gas in the particles can eventually escape from their hosts. For comparison, we have also adopted two different setups in other runs, one with a gas return fraction of $50\%$, the other with the gas return fully disabled. The return of the remaining gas in a particle begins 2.0 Myr after star formation, based on previous simulations (e.g. \citealt{Dale2014, Kim2018, li2019disruption, 2021MNRAS.506.5512F}). Moreover, the gas stored in a particle will return to the surrounding grid cells as an exponential function of time on a time scale $\tau_{rtn}$
 \begin{equation}
     \dot M_{gas} = -\frac{M_{gas}}{\tau_{\rm{rtn}}}. 
 \end{equation}
 Here in our simulation, a time scale of $\tau_{\rm{rtn}} = 5.0 Myr$ is adopted. At each time step, gas leaked from a particle is injected into grid cells within the particle's control volume with an initial velocity equal to that of the host particle, and subsequently accelerated by stellar winds and radiation pressure applied to the same region.

\subsubsection{Radiation}
\label{sec:ra}
Multiple works have shown that radiation pressure and photoionization dominate over gas pressure in the central region of molecular clouds, acting as the main mechanisms for cloud destruction and disrupting star formation (\citealt{Ceverino2013, Sales2014}). More recently, radiation feedback has been proven capable of removing gas from the center of a molecular cloud and carving channels into the outer region, allowing supernova remnants to pass through without much resistance (\citealt{Dale2014, Fujii2021}). \cite{Emerick2018} shows that these two mechanisms can play a catalytic role in the launch of galactic outflow in a low-mass dwarf galaxy. To investigate the effect of radiation, we incorporate both radiation pressure and radiation heating feedback in our simulation. Note that photoionization is not explicitly included in our simulation. Instead, we employ photoelectric heating in our cooling/heating source term for the following reasons: 1) resolving the detailed chemical evolution is not the main intention of this work, and 2) most of the gas in M82 remains neutral throughout the starburst, while ionized gas in the dense area will recombine quickly. From a statistical point of view, photoionization can be treated as a heating term. In our simulation, only UV radiation is considered, as it carries the most energy and momentum at the early age of a starburst galaxy. We use two separate modules to account for the effect of the photoheating and radiation pressure separately, both of which will be described in this section.

For heating rate calculation, we use the built-in photoelectric heating module of GRACKLE, which is based on Equation 1 in \cite{Wolfire1995}. To compute photoelectric heating, the intensity of the interstellar radiation field must first be determined. To this end, we use an approach different from the optically thin assumption, considering that most parts of the M82's gas disk, especially the star-forming clumps and SNR bubble shells, are opaque to UV radiation (\citealt{Murray2009}). \cite{Thompson2009} suggested that the characteristic opacity of UV radiation is $\kappa_{UV} \sim 10^{3}\, \rm{cm}^{2} \cdot \rm{g}^{-1}$ in the neutral gas. The two-temperature Planck mean opacity calculated from OPTAB (\citealt{Hirose2021}) also shows that for a typical OB star with an effective surface temperature of more than $10^{5}$ K surrounded by ambient gas with a temperature ranging from 10 to 20000 K, the opacity can be as high as $10^{2} \sim 10^{3}\, \rm{cm}^{2} \cdot \rm{g}^{-1}$. In such an environment, the inverse-square law could overestimate the intensity of the UV radiation field by a few orders of magnitude in a region far away from the UV source. To correct for this effect, we instead use a modified inverse-square law to estimate the intensity of the radiation field. More specifically, the UV flux distribution is obtained by solving the point-source divergence equation with absorption under spherical symmetry as follows.
\begin{equation}
    \nabla F = -F \kappa \rho, 
\end{equation}
where $\kappa $ is the opacity of the medium. The radial fraction of the equation gives
\begin{equation}
    \frac{1}{r^{2}} \frac{\partial r^{2} F}{\partial r} = -F \kappa \rho . 
\end{equation}
Consequently, the profile of the radiation field resulting from a point source is 
\begin{equation}
    F_{r} = F_{0} \frac{1}{r^{2}} e^{-\kappa \rho r}, 
\end{equation}
or alternatively 
\begin{equation}
    F_{r} = F_{0} \frac{1}{r^{2}} e^{-r/\lambda_{UV}},
    \label{eq:UV_intensity}
\end{equation}
where $\lambda_{UV}=\kappa \rho $ is the local UV attenuation length. In practice, to reduce computational cost, $\lambda_{UV}$ is calculated using the average density in each mesh block ($8\times 8 \times 8$ gas cells around the star particle). At each time step of the simulation, the UV radiation intensity at a gas cell is calculated by the summation of UV radiation fields produced by star particles within the distance of 5 times $\lambda_{UV}$. A maximum distance of 200 pc is also enforced when performing the summation.

On the other hand, radiation pressure is implemented as a momentum source. \cite{Sales2014} suggested that the radiation pressure is capable of accelerating the gas to a velocity up to $ 50 \sim 100 \rm{km\,s^{-1}}$ especially in high-density regions with $n_{H} > 1000 \rm{cm^{-3}}$, where the recombination time is short. Given the very high infrared (IR) opacity in molecular clouds, IR photons reradiated from dust grains will scatter multiple times before escaping, providing an extra boost to radiation pressure (\citealt{Agertz2012, Sales2014}). In order to resolve the radiation pressure from multiple sources while keeping a reasonable computation cost, we reuse the UV field derived previously in equation \ref{eq:UV_intensity} to calculate the local volumetric UV absorption rate
\begin{equation}
    I = F \cdot \kappa \cdot \rho
\end{equation}
The volumetric momentum injection rate of radiation pressure is then given by
\begin{equation}
    \dot P_{\rm{rad}} = (\eta_{1}+\eta_{2}\tau_{\rm{IR}})\frac{I}{c}.
    \label{eq:rad_p}
\end{equation}
where $\eta_{1}$ and $\eta_{2}$ adopt the value of $1.0$ in areas that are fully opaque to UV and IR. Optical depth $\tau$ takes the simplified form of
\begin{equation}
    \tau = \kappa \cdot \rho \cdot L
\end{equation}
where $L$ is the length of the transmission path of the light. The IR optical depth $\tau_{\rm{IR}}$ in M82's star cluster is assumed to be $10 \sim 100$ based on the observed densities (\citealt{Agertz2012}), therefore IR may be the dominant term in radiation pressure. In the calculation of optical depth, we use a similar approach to that in the calculation of self-shielding in GRACKLE (\citealt{Smith2016}), we use the local Jeans length $R_{J}$ as the radius of the host clump within which the grid cell resides, therefore
\begin{equation}
    \tau = \kappa \cdot \rho \cdot R_{J}
\end{equation}
Finally, we compute the radiation pressure exerted on the boundary of a grid cell by dividing the volumetric momentum injection rate by the surface area of the grid cell. Since we treat radiation pressure as a momentum source term, the effect of radiation pressure is then
\begin{equation}
    \frac{d P}{d t} = -\nabla P_{\rm{Radiation}}
\end{equation}

 \subsubsection{Stellar Wind}
In addition to radiation and SNe, the stellar wind is a non-negligible source of feedback, which could be comparable to that of SNe under certain circumstances (\citealt{Agertz2012}).
We adopt a metallicity-dependent model proposed by \cite{Jorick2021} to calculate the stellar wind's mass loss rate and final velocity.

\begin{table*}
\centering
\caption{List of all the simulations performed, the first and second columns are the name and mass of gas disc, respectively. The third, fourth and fifth column indicate the setting of supernovae, stellar winds and radiation feedback, respectively. The sixth and seventh columns indicate the gas return fraction and initial gas metallicity respectively (see section \ref{sec:Simulation} for more details). Simulation FD is specified as the fiducial run of our work. vSN (S16) stands for the variable SNe feedback model given by \protect\cite{Sukhbold2015}.} 
    \label{tab: Simulations}
    \begin{tabular}{|l|l|l|l|l|l|l|}
         \\[-2ex]\hline 
         \hline \\[-2ex]Simulation name & $\rm{M_{gd}}$ & supernovae & Stellar Wind & Radiation & Gas Return & $\rm{Z_{ini}}$\\        
         \\[-2ex]\hline 
         \hline \\
         FD & $3.0 \times 10^{9} \rm{M_{\odot}}$ & vSN(S16) & Yes & Yes  & $100\%$ & $0.02 Z_{\odot}$ \\
         fSN & $3.0 \times 10^{9} \rm{M_{\odot}}$ &$\rm{E_{SN}} = 10^{51} \rm{erg}$ & Yes & Yes & $100\%$ & $0.02 Z_{\odot}$\\
         nSN & $3.0 \times 10^{9} \rm{M_{\odot}}$ & No &Yes & Yes & $100\%$ & $0.02 Z_{\odot}$\\
         nWind & $3.0 \times 10^{9} \rm{M_{\odot}}$ & vSN(S16) & No & Yes & $100\%$ & $0.02 Z_{\odot}$\\
         nRad &  $3.0 \times 10^{9} \rm{M_{\odot}}$ & vSN(S16) & Yes & No & $100\%$ & $0.02 Z_{\odot}$\\
         nFB & $3.0 \times 10^{9} \rm{M_{\odot}}$ & No & NO & No & $100\%$ & $0.02 Z_{\odot}$\\
         GL00 & $3.0 \times 10^{9} \rm{M_{\odot}}$ & vSN(S16) & Yes & Yes & $0\%$ & $0.02 Z_{\odot}$\\
         GL05 & $3.0 \times 10^{9} \rm{M_{\odot}}$ & vSN(S16) & Yes & Yes & $50\%$ & $0.02 Z_{\odot}$\\
         M25 & $2.5 \times 10^{9} \rm{M_{\odot}}$ & vSN(S16) & Yes & Yes & $100\%$ & $0.02 Z_{\odot}$\\
         M35 & $3.5 \times 10^{9} \rm{M_{\odot}}$ & vSN(S16) & Yes & Yes & $100\%$ & $0.02 Z_{\odot}$\\
         Z01 & $3.0 \times 10^{9} \rm{M_{\odot}}$ & vSN(S16) & Yes & Yes & $100\%$ & $0.1 Z_{\odot}$\\
         Z05 & $3.0 \times 10^{9} \rm{M_{\odot}}$ & vSN(S16) & Yes & Yes & $100\%$ & $0.5 Z_{\odot}$\\         
         \hline \\
    \end{tabular}
\end{table*}

\subsubsection{Core-collapse SNe}
 Conventionally, stars within the zero-age main sequence mass range of $8\sim 30 M_{\odot}$ are thought to end their life through the outbreak of core-collapse supernovae (CCSN). Most previous models use a fixed thermal energy injection value of $10^{51}$ erg. However, recent studies indicate that the energy and mass of the ejecta as well as the composition of the ejecta depend on certain parameters, especially the ZAMS mass of the progenitor. In our study, we mainly follow the work of \cite{Thales2021}. We divide the total mass of each star particle (star cluster) into multiple ZAMS stars according to a \cite{Kroupa01} IMF. In practice, instead of creating many refined small particles, we store a list of individual virtual stars' information, including mass and expected lifetime, for each star particle (star cluster). At each time step, we update the age of virtual stars and loop over the lists to locate all the massive stars that have exceeded their expected lifetime and then trigger either a CCSN or failed SN for each such virtual star. The injected energy, ejecta mass, and metal of each SN (failed SN) are then calculated from the \cite{Sukhbold2015} CCSN model accordingly. Stellar wind and UV radiation before the outbreak of CCSN have already cleared the path for SNe \cite{Dale2014}. Therefore, we directly inject 100\% of SNe's mass, metal, and energy output, together with the much enhanced stellar wind produced in the last few years of the pre-supernova stage (\citealt{Sanskriti2017}), into grid cells within the control volume at the very end of their lifetime.

The lifetime of massive stars is calculated using a scheme similar to that of \cite{Thales2021}, which involves linearly interpolating the table provided by \cite{Portinari1998} with an assumed metallicity equals to the initial gas metallicity, i.e., $Z = 0.02 Z_{\odot}$ for most runs. This model has been proven robust with small discrepancies compared to other stellar lifetime models such as \cite{Schaller1992} and \cite{Paxton2010}. Note that not all stars with a ZAMS mass above 8 $M_{\odot}$ will end their lives as CCSN. A noticeable fraction of OB stars may collapse directly to form black holes during the core collapse, which is commonly referred to as failed SNe (\citealt{1986ARA&A..24..205W, 1999ApJ...524..262M}). Failed SNe often release considerably less energy, typically on the order of $10^{49}$ erg, and result in a much smaller amount of mass return (e.g. \citealt{Sukhbold2015}). In our simulation, we apply the same mass and energy injection scheme for CCSN to these failed SNe, but with their mass, metal, and energy injection rates inferred from the \cite{Sukhbold2015} model. 

Apart from the variable SNe model from \cite{Sukhbold2015}, we have also adopted the widely used fixed energy ($E_{\rm{SN}} = 10^{51} \rm{erg}$) SNe model in run fSN, and compare the effect of the two different setups in the following sections.

\subsection{radiative cooling and heating}
Radiative cooling and heating play an important role in the evolution of ISM. In our simulation, we use GRACKLE (\citealt{Smith2016}), an open-source chemistry and cooling library, to solve radiative cooling and heating. GRACKLE provides primordial and dust chemistry calculations, as well as tabulated metallicity-dependent cooling functions with and without self-shielding. Our simulation use tabulated cooling function calculated from CLOUDY, which account for collisional ionization and ionization from UVB.

The cooling rate of the gas highly depends on the metallicity. For a star-forming gas cloud with constant feedback and metal enrichment, the metallicity can vary from the initial value of 0.02 $Z_{\odot}$ to as high as 20 $Z_{\odot}$, causing the cooling rate to vary by an order of magnitude. This variation in the cooling rate could significantly affect the evolution of molecular clouds, the HII region, and the multiphase outflow. In order to obtain a more accurate result, we use a passive scalar as the tracer for metal elements in the ISM. Moreover, we include the SNe metal enrichment based on the isotope yield provided in the work of \cite{Sukhbold2015}. Therefore, we can trace the detailed evolution of the metal elements in the ISM and calculate an accurate cooling function accordingly.

\begin{figure*}
    \centering
    \includegraphics[width=1.65\columnwidth]{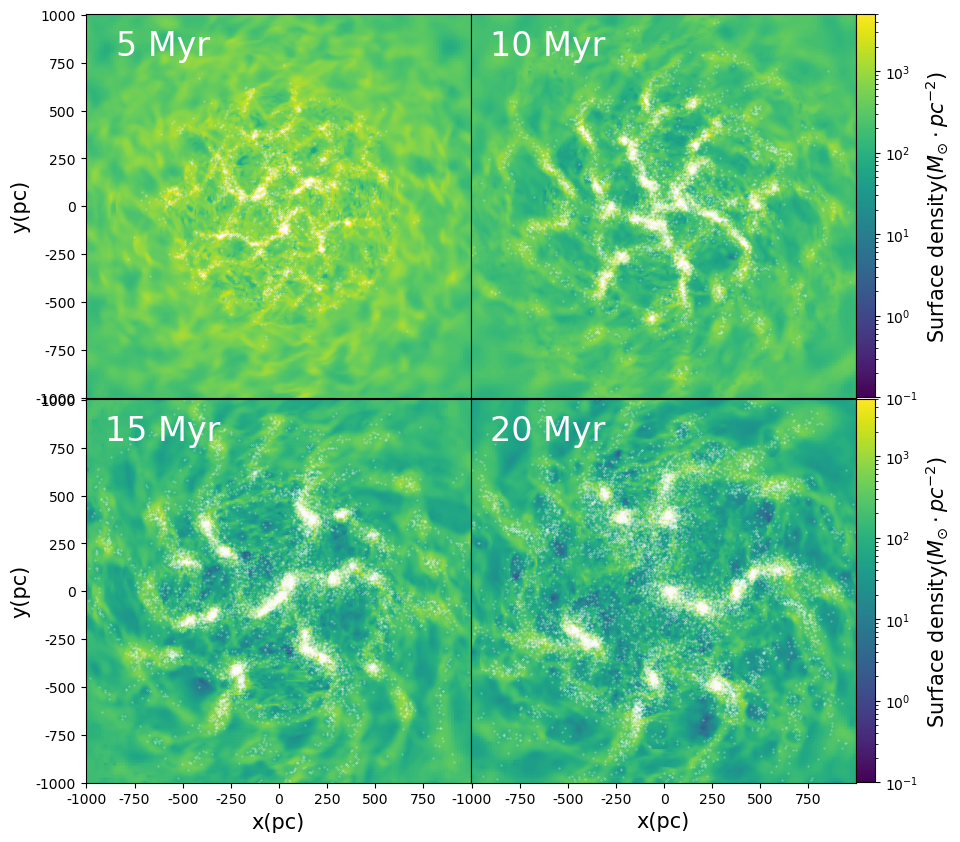}
    \caption{Face-on view of the gas disk in the FD simulation, showing the projected gas density overlaid with star particles (white dots). This picture shows star particles form predominantly within high-density clumps. The bottom panels illustrate how superbubbles develop around clustered star particles. In these panels, the large number of star particles partly obscures the underlying high-density clumps.}
    \label{fig:sf_process}
\end{figure*}

The gas disk of M82 is a highly gas-rich and dust-rich environment, where the optical-thin assumption may fail due to the large opacity in some regions. Previous studies argue that self-shielding can significantly reduce the cooling rate by an order of magnitude in cool to warm regions with relatively low density ($\sim 1 \rm{cm^{-3}}$), which is enough for the radiation feedback to become significant to support the clouds from collapsing (\citealt{Andrew2018}). In our simulation, the self-shielding option provided by GRACKLE is turned on, in which the self-shielding of both HI and HeI from the UV background (UVB) is included. The \cite{Haardt2012} cosmic UV background model is adopted to mimic the UVB. The detailed derivation of the tabulated self-shielding cooling function in GRACKLE is described in the work of \cite{Smith2016} and \cite{Rahmati2013}.

We also include photoelectric heating provided by GRACKLE in our simulation, detailed setup has been described in Section \ref{sec:ra}. Many previous simulations incorporate a delay cooling method, disabling radiation cooling around star clusters for a period of time, therefore ensuring that the SNR bubbles expand to the Sedov stage without losing much of their energy in the free expansion stage and increasing the effect of hot gas (e.g.\citealt{Agertz2012}). However, with UV radiation heating included, star clusters can efficiently ionize the gas within a Strömgren radius $r_{S}$.
\begin{equation}
    r_S = \left( \frac{3}{4 \pi} \cdot \frac{N_{\text{ion}}}{n_H^2} \right)^{1/3}
\end{equation}
in which $N_{\text{ion}}$ is the number of H ionizing photons produced by the OB stars in the star cluster. We note that in a typical SSC in M82, the Strömgren radius can reach $5 \sim 20 \rm{pc}$ in dense areas with $n_{H} \sim 1000 \,\rm{cm^{-3}}$, enough for the SNR to expand while conserving most of its energy. Overall, using photoelectric heating instead of delay gas cooling around star clusters provides a more sophisticated account for the mechanisms involved in the launch of the SNR bubbles while avoiding the introduction of artifacts.

\subsection{Simulations}
\label{sec:Simulation}
Our simulations were performed with the state-of-the-art grid-based MHD code Athena++ (\citealt{Stone2020}). In our simulation, we use the Harten–Lax–van Leer–Contact (HLLC) scheme as the Riemann solver. For self-gravity, we used the multi-grid self-gravity solver module developed by \cite{Tomida2023} to solve the self-gravity hydrodynamic equations. At present, our simulations do not incorporate magnetic fields. The simulations were run in a cubic box with a side length of 4 kpc and an outflow boundary condition. Static mesh refinement (SMR) is used to refine the cells in the central starburst region to up to 4 pc resolution. The detailed description of the SMR scheme is given in Appendix \ref{sec:smr}. Note that the gas disk is not truncated in our simulations. Although not shown in the manuscript, we performed additional test runs and confirmed that this choice does not affect our results. The detailed setups of the simulations that we performed are listed in Table \ref{tab: Simulations}.

\begin{figure}
    \centering
    \includegraphics[width=\columnwidth]{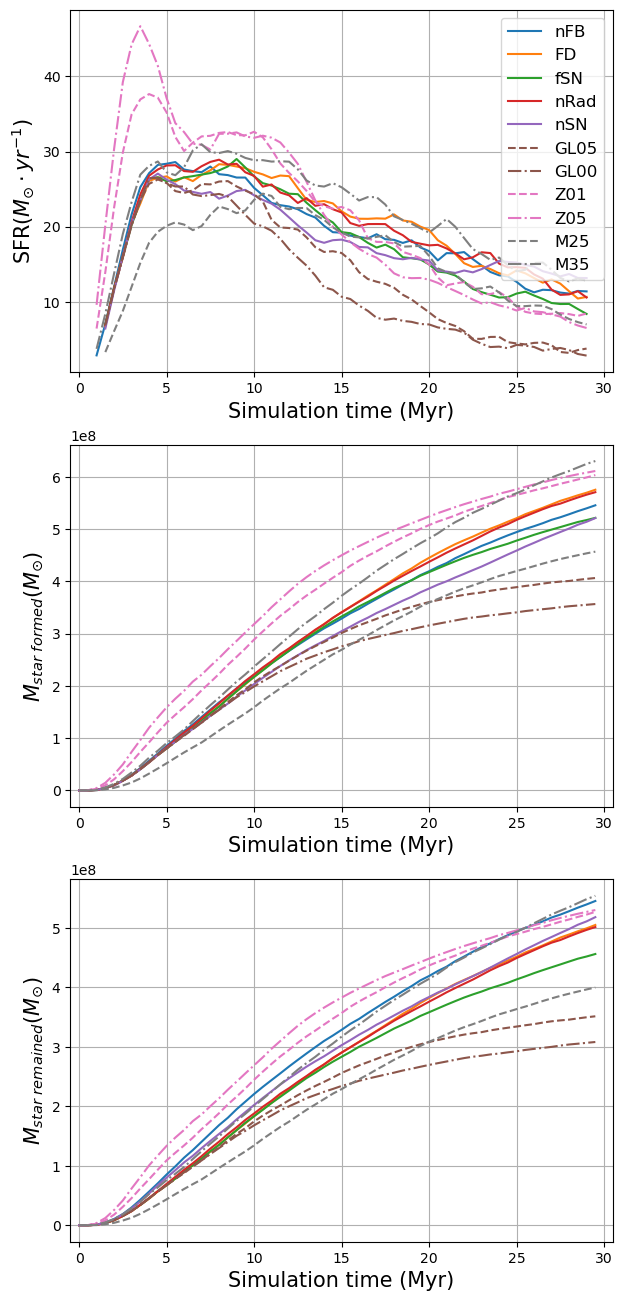}
    \caption{Top: The star formation rate as a function of time in various simulations. Middle: The cumulative mass of stars formed since the beginning of the simulation. Bottom: The remaining stellar mass within $r<1000$ pc that has yet undergone SNe as a function of time.}
    \label{fig:sfh}
\end{figure}

\section{Starburst in our simulations}
Our simulations can generate a starburst in the central region of the gas disc. In this section, we report the features of the starburst and the stars formed. We begin with the star formation history and efficiency, and then probe the properties of star clusters.  
\subsection{star formation history and
efficiency}
\label{sec:starburst1}
M82 has undergone multiple starburst events throughout its history (\citealt{mayya2009m82}). The majority of stars in the stellar disk are between 300 Myr and 1 Gyr. The recent starburst, which occurred within the past $15 \sim 50$ Myr in the nuclear region (within $r<625$ pc), is responsible for the current galactic outflow. Some studies suggest that this recent starburst comprised two distinct events (\citealt{1993ApJ...412...99R, Schreiber2003}), with a major burst peaking around $8 \sim 15$ Myr ago and having a peak SFR of $160\, \rm{M_{\odot}} \cdot \rm{yr}^{-1}$ and lasting for only about 1 Myr. 

Figure \ref{fig:sf_process} illustrates the star formation process in the FD simulation. At $t=5$ Myr, gravitational instability leads to the formation of numerous small and medium-sized gas clumps, particularly in the nuclear region. Concurrently, a significant number of star particles emerge. As time progresses, these gas clumps grow and spawn more star particles. Clumps merge to form larger clumps, and star particles would form super star clusters. By 15-20 Myr, a substantial number of stars and clusters have formed, leaving only a small volume occupied by dense gas. We find that most of the stars form within $r<1.0$ kpc, while hardly any stars form beyond $r>1.2$ kpc due to the low gas density in that region.

The top panel of Figure \ref{fig:sfh} quantitatively illustrates the star formation history (SFH) in our simulations. Most simulations exhibit a rapid increase in the star formation rate (SFR) from nearly zero at $t=2$ Myr to a peak of 20-30 $\rm{M_{\odot}\, yr^{-1}}$ at $t\sim 5$ Myr, followed by a gradual decline to around 4-12 $\rm{M_{\odot}\, yr^{-1}}$ at $t\sim 30$ Myr. Compared to \citealt{Schreiber2003}, our simulated starburst exhibits a flatter profile, a lower peak SFR, and a longer duration. The top panel of Figure \ref{fig:sfh} also highlights the impact of different feedback processes, gas return, initial gas disc mass, and initial metallicity on the SFH. 

Disabling any single stellar feedback mechanism has a minor effect on the SFH. However, disabling all three feedback processes will reduce the SFR by $5\sim 10 \%$, which seems somewhat abnormal. 
Meanwhile, switching to a fixed SNe energy model, which injects roughly twice the energy of the fiducial variable SNe model, moderately reduces the SFR after $t\sim 10 $ Myr. The probable reason is that stellar feedback with moderate strength can positively influence star formation. However, excessively strong feedback can suppress star formation activity. We conclude that the stellar feedback has a minor impact on the overall SFH of the starburst. This is likely because, on the starburst timescale of 30 Myr, the amount of gas that can collapse and how quickly the gas can collapse is mostly determined by the initial conditions of the gas disc. Additionally, the gas return process implemented in most simulations allows particles to release the accreted gas mass back to the ISM, effectively sustaining star formation activity at later time. 

Figure \ref{fig:sfh} shows that reducing the gas return fraction from $100\%$ to $50\%$ and $0\%$ significantly lowers both the SFR and overall SFE within 30 Myr, as a substantial fraction of gas becomes locked in sink/star particles. Meanwhile, varying the initial metallicity $\rm{Z_{init}}$ notably impacts the SFH. Higher initial metallicity leads to faster cooling, resulting in a more violent starburst during the first 15 Myr, with peak SFRs of $37\, \rm{M_{\odot} \,yr^{-1}}$ in the Z01 simulation and $45\, \rm{M_{\odot} \, yr^{-1}}$ in Z05. However, this intense early starburst is followed by rapid quenching after $t>15$ Myr, leading to a final SFE and total stellar masses similar to those of the fiducial run (FD). In addition, increasing $\rm{M_{gd}}$ will moderately enhance the SFR at $t>15$ Myr, while a lower $\rm{M_{gd}}$ will suppress star formation activity throughout the simulation. 

The middle panel of Figure \ref{fig:sfh} presents the total stellar mass formed as a function of time. At $t\sim 30$ Myr, the total stellar mass in most simulations ranges from $5.0\times 10^{8} \,\rm{M_{\odot}}$ to $ 6.0\times 10^{8} \,\rm{M_{\odot}}$, corresponding to a total input of SN energy of $2.3 \sim 2.7 \times 10^{57} \rm{erg}$ for the variable SN model and $\sim 4.7 \times 10^{57} \rm{erg}$ for the fixed SN model. Reducing the gas return fraction to $50\%$ or $0\%$, or adopting a lower gas disk mass of $\rm{M_{gd}}=2.5 \times 10^{9} \,\rm{M_{\odot}}$, results in a lower total stellar mass of $3.5\times 10^{8} \,\rm{M_{\odot}}$ to $ 4.5\times 10^{8} \,\rm{M_{\odot}}$. This implies that approximately $12\%-20\%$ of the initial gas disk mass was converted into stars over 30 Myr. Observational estimates of the total stellar mass formed during the past $\sim 15-30$ Myr range from $2.0\times 10^{8} \,\rm{M_{\odot}}$ to $ 3.5\times 10^{8} \,\rm{M_{\odot}}$ (\citealt{1993ApJ...412...99R,Schreiber2003,2009ApJ...697.2030S}), which is broadly consistent with our simulations with reduced gas return fractions or lower initial gas disk masses. However, observational estimates of the SFH are subject to uncertainties due to factors such as dust absorption, projection effects, and assumptions about the initial mass function and stellar evolution (e.g.\citealt{Schreiber2003}). 

Another way to compare our results with observational constraints is to examine the total mass of stars formed during the starburst that remain within the nuclear region. The bottom panel of Figure \ref{fig:sfh} shows the stellar mass within $r<1000$ pc as a function of time in our simulations, suggesting that $\sim 20\%$ of the formed stellar mass has either undergone SN explosions or escaped from the nuclear region. At $t=30$ Myr, the total remaining stellar mass within $r<1000$ pc ranges from $3.0\times 10^{8} M_{\odot}$ to $6.0\times 10^{8} M_{\odot}$, which is below the stellar mass inferred from observations in the nuclear region (within $\sim500-600$ pc), as reported by \cite{Forster2001}. 

\begin{figure}
    \centering
    \includegraphics[width=\columnwidth]{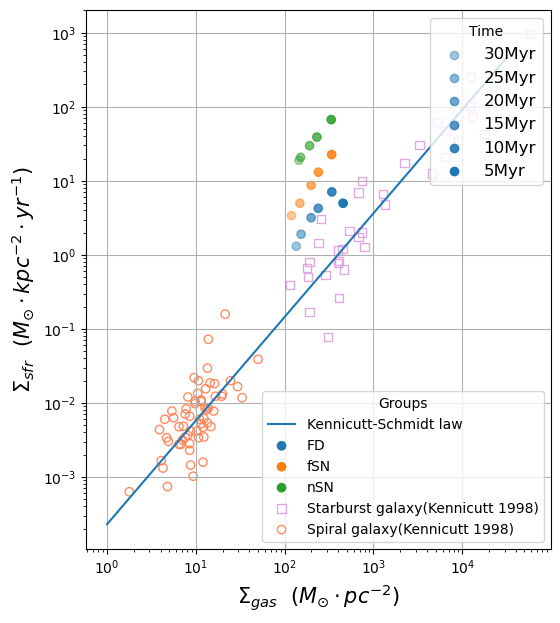}
    \caption{The surface star formation rate density against the surface gas density in the FD (filled blue circle), fSN (filled yellow circle) and nSN (filled green circle) simulations, compared to the to the Kennicutt-Schmidt law (solid blue line), starburst galaxies and spiral galaxies in \citealt{1998ApJ...498..541K}. For clarity, the fSN and nSN data points have been shifted upward by 0.5 and 1.0 dex, respectively.} 
    \label{fig:kennicutt}
\end{figure}

In addition, we explore the relation between the surface SFR density, $\Sigma_{SFR}$, and cold gas surface density, $\Sigma_{gas}$, in our simulations to assess the validity of our star formation and feedback models. As shown in Figure \ref{fig:kennicutt}, the $\Sigma_{SFR}$ at a fix $\Sigma_{gas}$ in our simulations is 6 to 8 times higher than predicted by the classical Kennicutt-Schmidt law, but aligns well with the $\Sigma_{SFR}$-$\Sigma_{gas}$ relation observed in some of the starburst galaxies with highly efficient star formation activity. This elevated star formation activity may be attributed to several factors, including the excessively high surface density and the overall collapse or contraction of the nuclear region.

\begin{figure}
    \centering
    \includegraphics[width=\columnwidth]{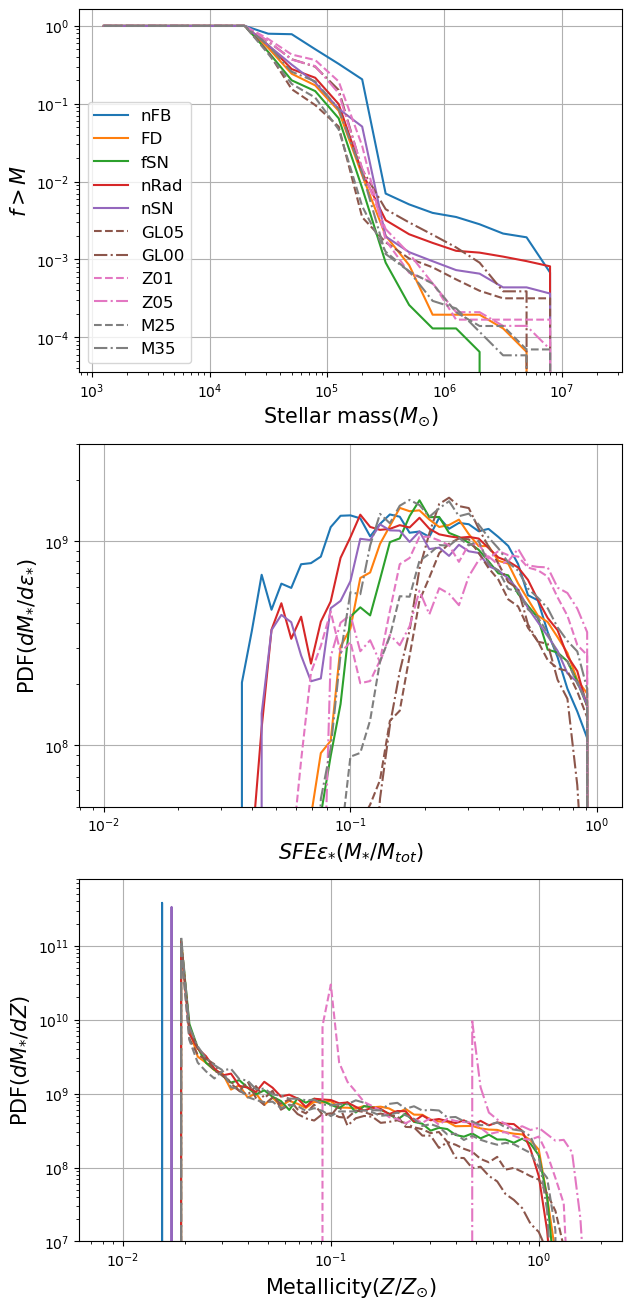}
    \caption{Top: the cumulative star cluster mass function in various simulations. Middle: the distribution of integrated star formation efficiency in sink/star particles. Bottom: The distribution of stellar metallicity in various simulations. The metallicity in nFB and nSN are $0.02\rm{Z_{\odot}}$, which has been shifted leftward slightly in the plot for the sake of clarity.}
    \label{fig:ccmf}
\end{figure}

\subsection{properties of the star clusters}
The properties of star clusters formed during the starburst can significantly influence the launch and development of galactic outflows. Previous studies have employed super star clusters (SSCs) to drive galactic winds in M82-like galaxies (\citealt{2013MNRAS.430.3235M, 2020ApJ...895...43S,2024ApJ...966...37S}). 
To investigate the properties of star clusters formed in our simulations, we utilize the friends-of-friends (FoF) algorithm to identify clusters and analyze their spatial distribution, metallicity, mass function (CMF), and other characteristics. Considering the limited spatial resolution, here we used a linking length $l_{\rm{fof}} = 8.0 \,\rm{pc}$.

The top panel of Figure \ref{fig:ccmf} presents the cumulative CMF in our simulations. Recent JWST observations by \cite{Levy2024} suggest a power-law CMF for star clusters in M82, $y\propto \alpha \rm{log}(x)$,  with a slope of $\alpha = 0.9 \pm 0.2$. Our simulated CMF exhibits a more bottom-heavy distribution. This discrepancy may be attributed to the imposed mass threshold, $M_{cnv}$, when converting a sink particle to a star particle. The higher-mass end of the CMF, less affected by this threshold, aligns better with the observed power-law slope of $\alpha = 0.9$.

While Section \ref{sec:starburst1} demonstrates that stellar feedback has a limited impact on the overall SFH, it significantly influences the cumulative CMF. Simulations nRad and nFB exhibit a more top-heavy CMF compared to other simulations, suggesting that stellar feedback effectively disrupts gas accretion in giant star-forming clumps, redistributes gas, and inhibits the formation of massive star clusters. Among the three feedback mechanisms, radiation feedback appears to be the most effective in suppressing the formation of SSCs, reducing the fraction of SSCs with masses exceeding $M_{*} > 10^{6} \,\rm{M_{\odot}}$ by $90\%$, and eliminating those with masses above $M_{*} > 10^{7}\, \rm{M_{\odot}}$. In contrast, the default variable SNe feedback mode has a more moderate impact on the formation of massive SSCs. 

On the other hand, all three stellar feedback mechanisms can suppress the integrated star formation efficiency within individual sink particles. The middle panel in \ref{fig:ccmf} shows the integrated SFE of sink particles across various simulations, which follows a roughly log-normal pattern. The median integrated SFE ranges from $10\%$ to $30\%$, which is generally consistent with previous simulations (\citealt{Dale2014,li2019disruption,2021MNRAS.506.5512F}), but higher than observational estimates of (\citealt{2020MNRAS.493.2872C}). This discrepancy may be partly attributed to our assumption of a $20\%$ star formation efficiency per free-fall time for all sink particles, regardless of their properties.

Compared to the FD simulation, disabling either radiation feedback or SNe feedback results in a higher fraction of star clusters with integrated star formation efficiencies below 0.1. This suggests that sink particles in these simulations accrete gas at a faster rate than in the FD run, accumulating 5-10 times more gas mass before being converted to star particles. These findings underscore the importance of radiation feedback and SNe feedback in suppressing gas accretion onto star-forming clumps.

The metallicity of newly formed stars in the nuclear region provides insights into the properties of the starburst and the impact of SN feedback. The bottom panel of Figure \ref{fig:ccmf} shows the distribution of stellar metallicities. In simulations without SN feedback, star clusters often retain their initial metallicity. However, in simulations with SN feedback, a significant fraction of star clusters experience metal enrichment, with some reaching metallicities several to tens of times higher than the initial value. In some cases, stars can even attain solar metallicity. Observations by \cite{2004ApJ...606..862O} indicate that most cool stars in M82's nuclear region have metallicities between $\frac{1}{10}$ and one Solar metallicity. Their analysis suggests a solar iron abundance and $\alpha$-element enhancement for these stars. In our simulations, the median stellar metallicity with core-collapse SN feedback is slightly lower than observed values for an initial gas metallicity of $\rm{Z_{ini}}=0.02\, \rm{Z_{\odot}}$, but becomes comparable for $\rm{Z_{ini}}=0.10 \,\rm{Z_{\odot}}$.

\section{Outflow}
Understanding the evolution of different gas phases during the launch of galactic outflows and the detailed structure of these outflows is crucial for gaining insights into the underlying physics and their impact on the ISM and CGM. In this section, we will examine how gas in various phases is entrained into the outflow and qualitatively assess the properties of the galactic outflow in our simulations. We will also investigate the effects of different feedback mechanisms, gas return fractions, and initial gas disk masses on the development of the outflow.

\subsection{Launch of the kpc scale multiphase wind}

\begin{figure*}
    \centering
    \includegraphics[width=0.75\textwidth]{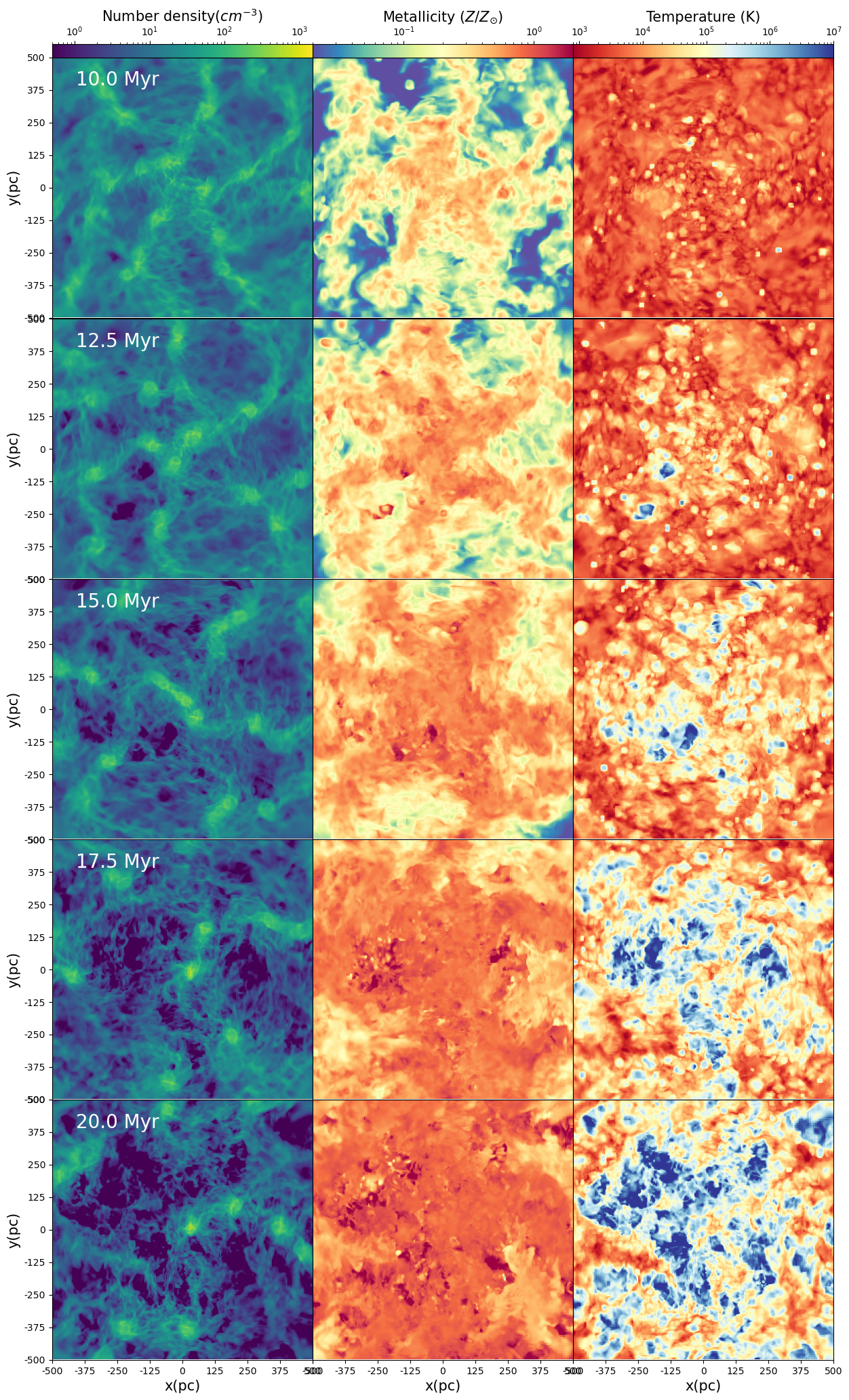}
    \caption{Face-on view of the central region of simulation FD. The left, middle, and right columns show the gas density, metallicity, and temperature. From the top to the bottom row, results at time t=10, 12.5, 15.0, 17.5, and 20.0 Myr are shown.}
    \label{fig:fd_zoomin_face}
\end{figure*}

\begin{figure*}
    \centering
    \includegraphics[width=0.99\textwidth]{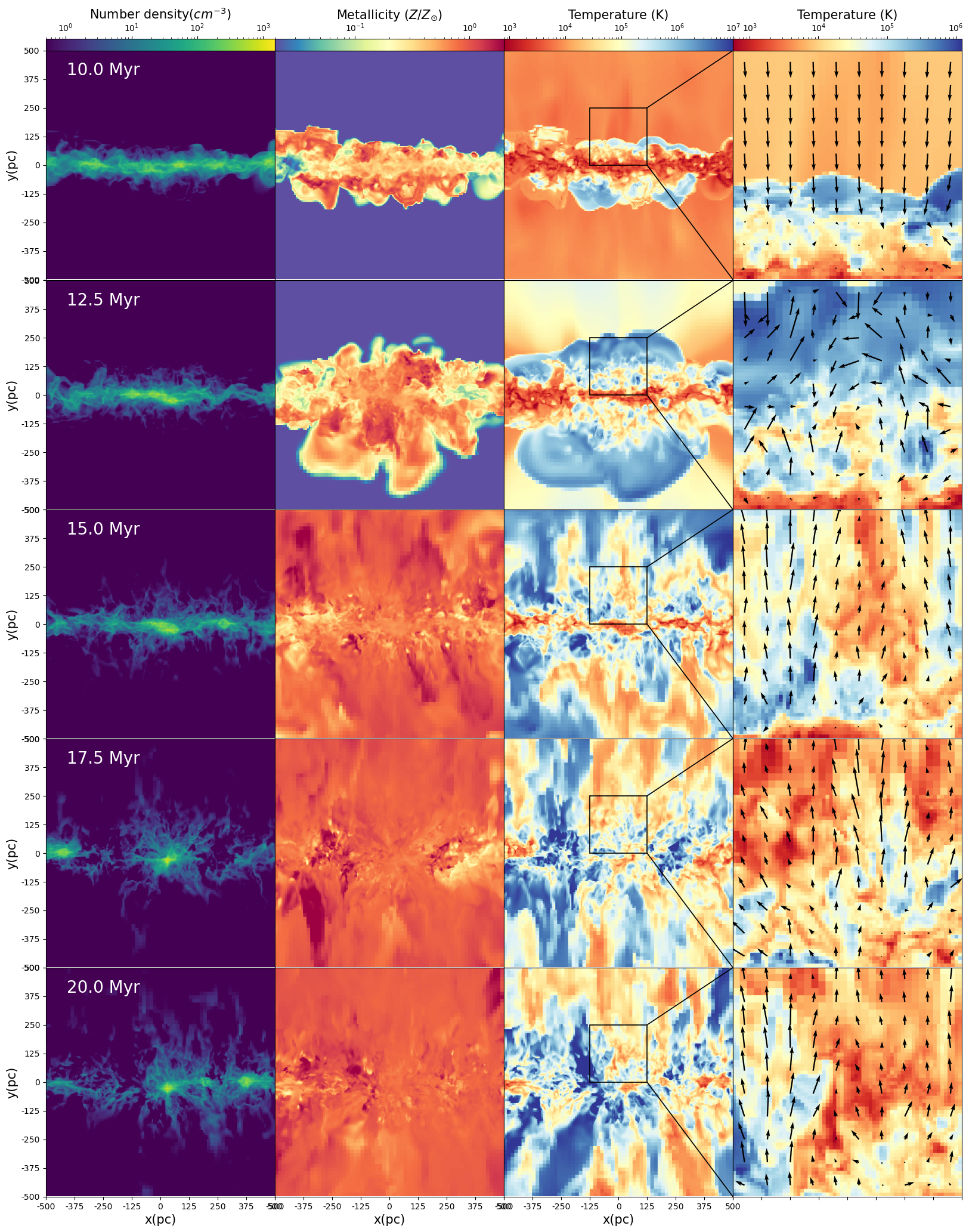}
    \caption{Edge-on view of the central region of simulation FD. The first, second, and third (from left to right) columns show the gas density, metallicity, and temperature, while the last column is the zoomed-in view of the temperature map overlapped with the velocity field}. From the top to the bottom row, results at time t=10, 12.5, 15.0, 17.5, and 20.0 Myr are shown.
    \label{fig:fd_zoomin}
\end{figure*}

\begin{figure*}
    \centering
    \includegraphics[width=1.7\columnwidth]{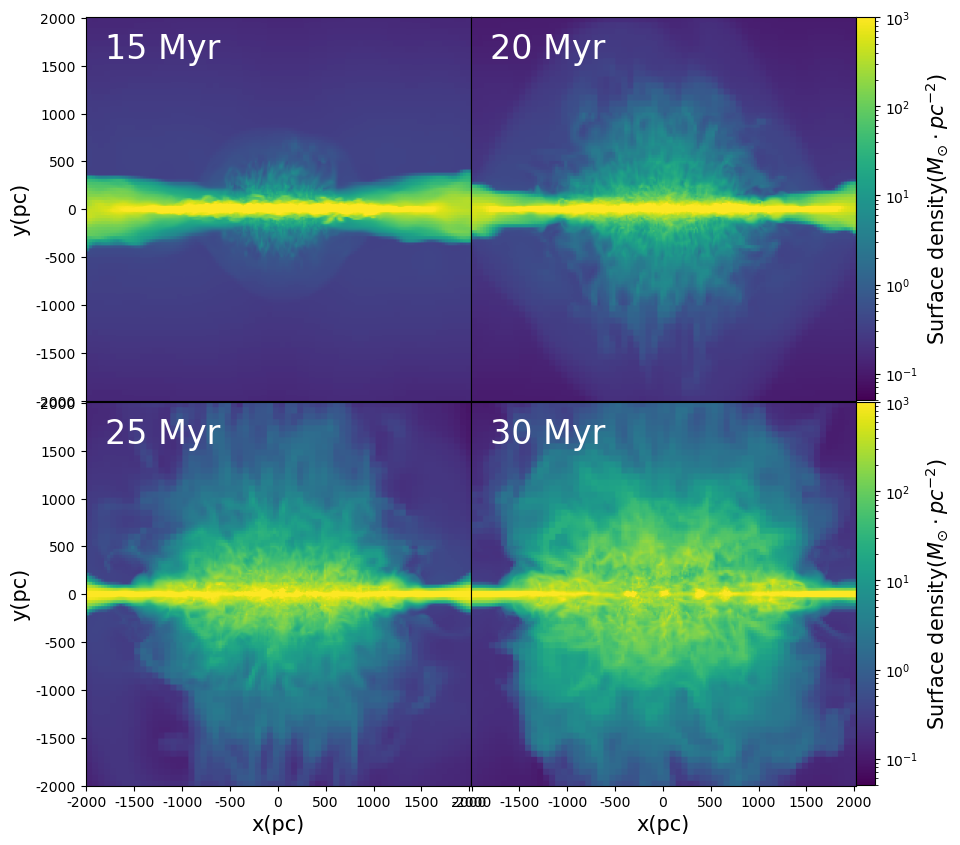}
    \caption{Edge-on view of the projected gas density in simulation FD. The four images show how galactic wind is launched in the widespread area across the galactic disk, eventually forming an outflow structure stretching to the edge of the simulated zone.}
    \label{fig:launch_all}
\end{figure*}

Generally, the launch of galactic winds in our simulations occurs in two stages: the formation and breakout of a superbubble foam, followed by the development of a kpc-scale multiphase wind. For the first stage, Figures \ref{fig:fd_zoomin_face} and \ref{fig:fd_zoomin} provide detailed insights into the launch of the multiphase outflow in the central starburst region of the FD simulation. As for the second stage, Figure \ref{fig:launch_all} presents an edge-on view of the whole simulation zone, showcasing the formation of the kpc-scale outflow.

The first stage, spanning from $t\sim 5$ Myr to $t\sim 15$ Myr, involves the formation and breakout of a superbubble foam. Initially, the turbulent ISM is dominated by cool gas residing in lower-density ($n \lesssim 5\ \rm{cm}^{-3}$) voids, intermediate to high density filaments, and very dense giant clouds with size of tens pc to 100 pc. A few isolated supernova remnants (SNRs) emerge around 5 Myr, expanding into the cool ISM. These SNRs form within dense giant clouds, leading to significant metal enrichment of the surrounding cool gas, as shown by the panels in the second and third columns of Figure \ref{fig:fd_zoomin_face}. The lower-density ISM in the voids is more easily ionized and expelled outward. As more SNe occur, their remnants merge to form a superbubble foam. In contrast, gas in intermediate to high density filaments and dense giant clouds are less affected by SNe feedback. A considerable fraction of the filaments can survive from the feedback at $t<15$ Myr, as shown by the density plots in Figures \ref{fig:fd_zoomin_face} and \ref{fig:fd_zoomin}. Between $t=10$ and 12.5 Myr, a single superbubble is formed and dominates the central starburst region, filled with warm and hot gas, numerous intermediate to high density cool filaments, and a few cool dense giant clouds (see the density and temperature panels in Figures \ref{fig:fd_zoomin_face} and \ref{fig:fd_zoomin}). This superbubble expands and eventually breaks through the dense gas disk before $t\sim15$ Myr, with outflowing gas reaching heights of 500 pc. 

\begin{figure*}
    \centering
    \includegraphics[width=1.8\columnwidth]{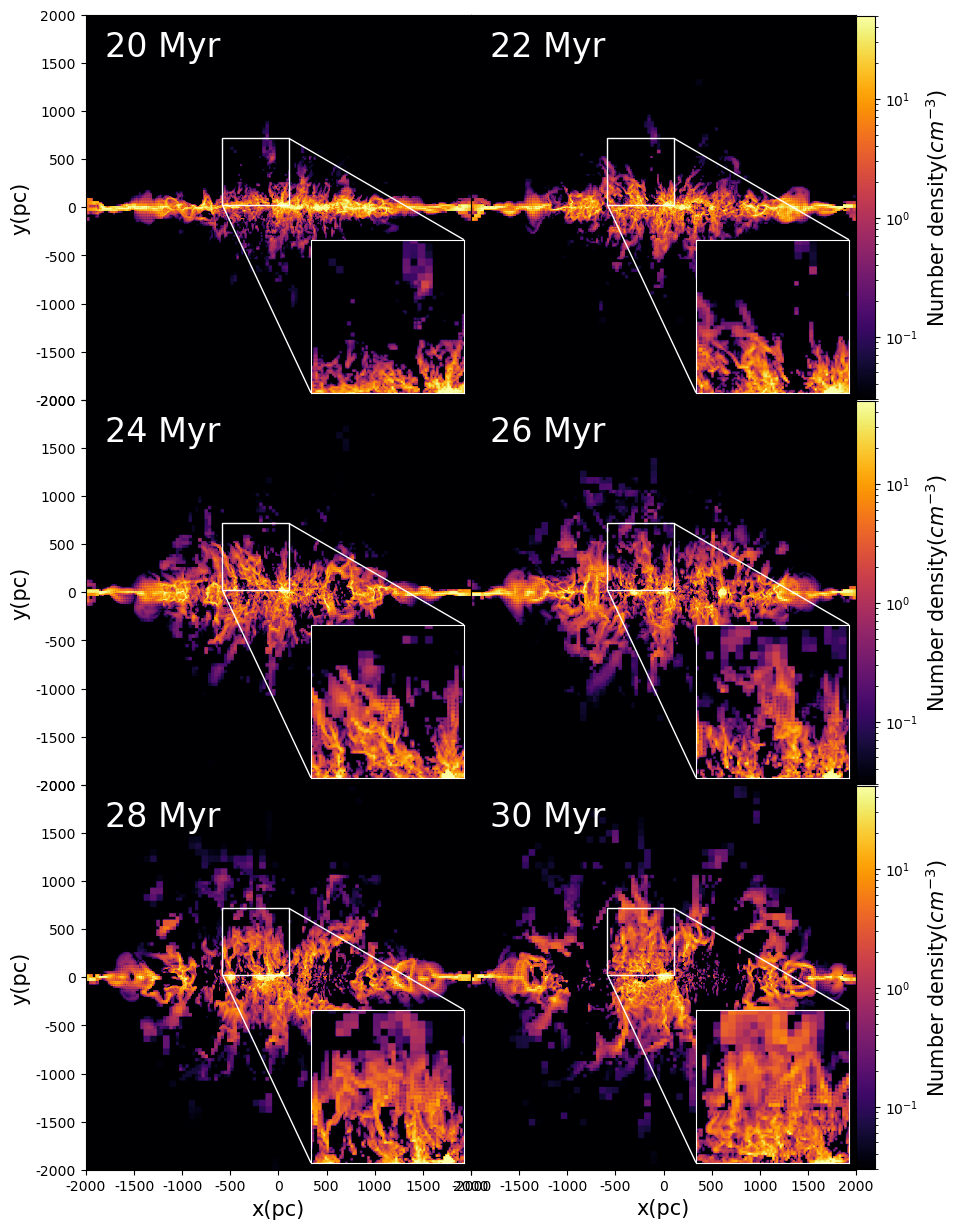}
    \caption{Filaments and cores of the neutral gas ($\rm{T} < 8000 \rm{K} $) in a slice of thickness 200 pc between $t=20$ and $30$ Myr in the FD simulation. The plug-in plot in each panel presents a zoom-in view of the interface between disc and outflow region with a size of 750$\times$750 pc.}
    \label{fig:fila_neutral}
\end{figure*}

\begin{figure*}
    \centering
    \includegraphics[width=1.8\columnwidth]{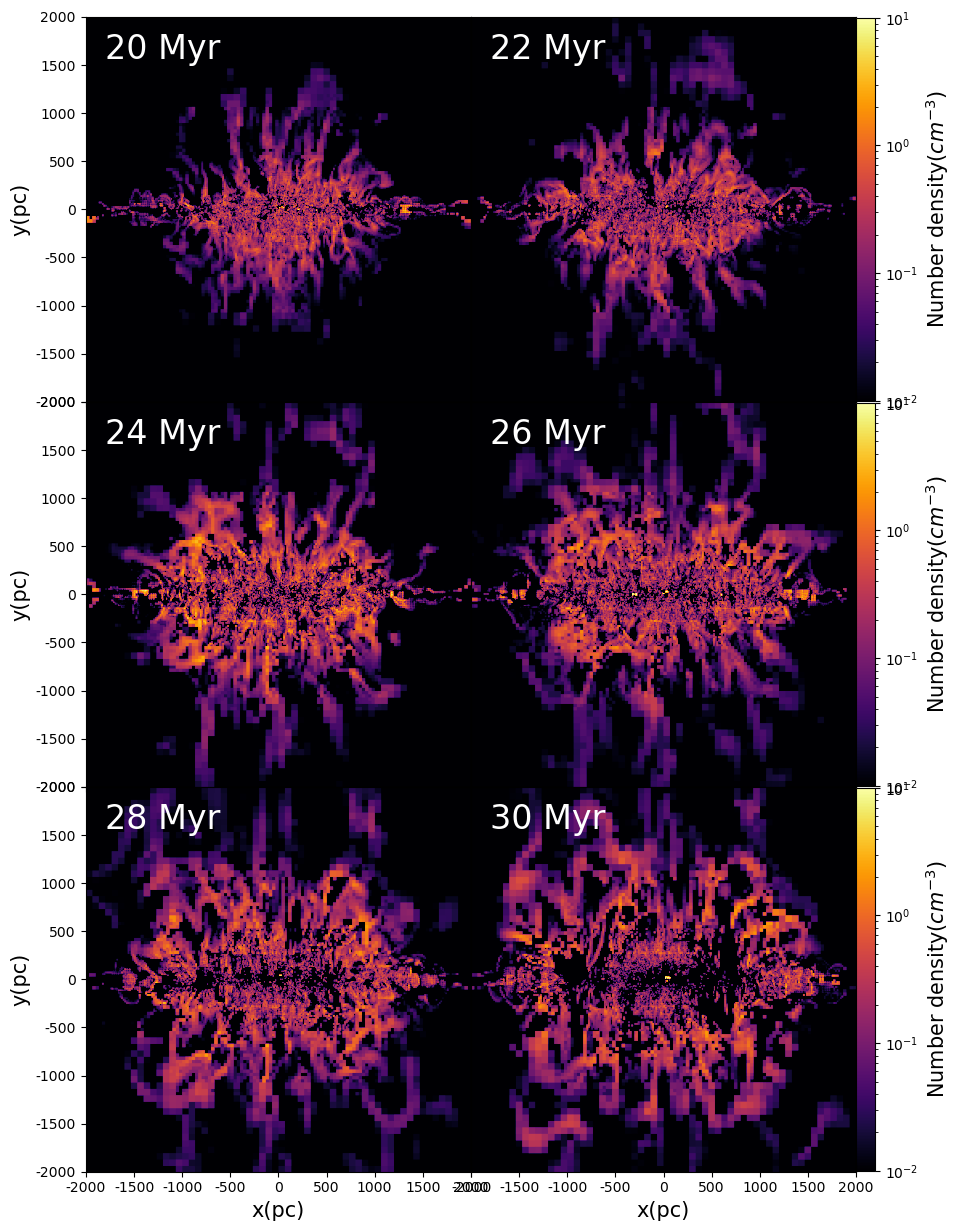}
    \caption{Similar to Figure \ref{fig:fila_neutral}, but for $\rm{H\alpha}$ emitting gas($8000 \rm{K} < \rm{T} < 23000 \rm{K} $).} 
    \label{fig:fila_ha}
\end{figure*}

\begin{figure}
    \centering
    \includegraphics[width=1.0\linewidth]{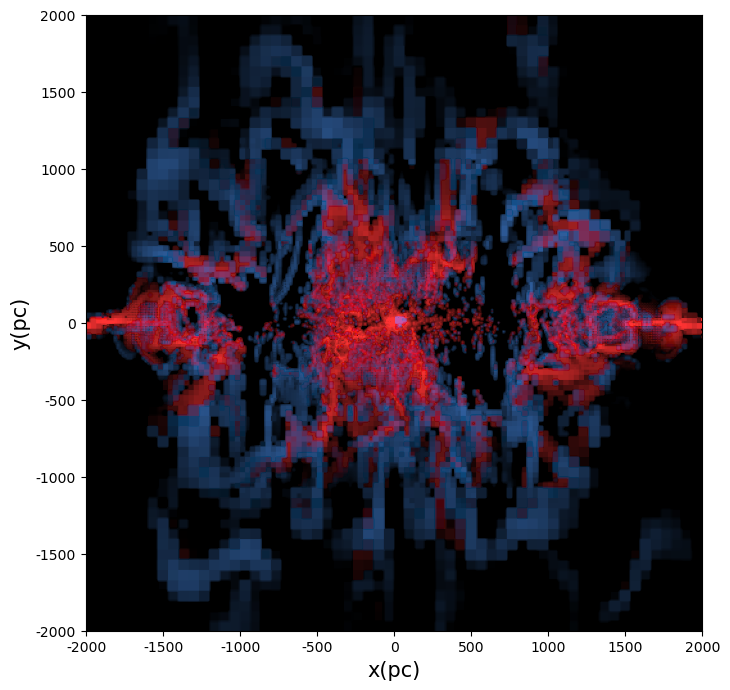}
    \caption{Shown is the overlapped image of the neutral filaments(red) and $\rm{H\alpha}$ emitting filaments(blue), both identified by eigenvalues of the Hessian matrix of gas density, at 30 Myr}.
    \label{fig:overlap}
\end{figure}


Following the superbubble breakout, hot gas moves outward along the minor axis of the disk at the highest velocity, followed by warm gas and, subsequently, cool gas, which is primarily carried in numerous filaments. The density and temperature panels in Figures \ref{fig:fd_zoomin_face} and \ref{fig:fd_zoomin}, especially the fourth column of Figure \ref{fig:fd_zoomin}, which provides a zoomed-in view of the interface between the central starburst region in the disc and the outflow within a $250 \times 250$ pc box, demonstrate that a significant fraction of cool gas in the disk region ($z < 125$ pc), largely contained in intermediate to high density filaments, can survive early feedback ($t < 10$ Myr) and subsequently be entrained into the wind at $t > 12.5$ Myr. The zoom-in panels at $t \gtrsim 12.5$ Myr in Figure \ref{fig:fd_zoomin} capture a considerable number of cool filaments still residing in the disk ($z < 125$ pc) as they begin to move outward.

Meanwhile, the zoom-in velocity field reveals substantial differences between the hot, warm, and cool phases flowing out of the disk, which can trigger mixing and Kelvin–Helmholtz instabilities. In this environment, interactions between the hot and cool phases can facilitate significant momentum transfer, accelerating the cool gas in filaments as it moves outward \citep{2018MNRAS.480L.111G,2020ApJ...895...43S,2022ApJ...924...82F}. During this process, the cool filaments can further stretched by the hot wind, their lengths increase, reaching hundreds of parsecs by $t=15$ Myr. These filament features in the wind are remarkably similar to recent JWST observations (\citealt{2024ApJ...967...63B}). 

The visual appearance of these cool filaments in the outflow resembles previous simulations of M82-like galaxies (e.g., \citealt{2008ApJ...674..157C, 2013MNRAS.430.3235M,2020ApJ...895...43S}). However, in our simulation, the dominant net contributor to the cool filaments in the outflow is the initially filamentary cool ISM in the disc. This contrasts with several previous studies, which typically attributed filament formation to interactions between initially non-filamentary, cold, clumpy clouds and the hot wind. For instance, simulations by \cite{2008ApJ...674..157C} show that small, cold clouds, produced by the disruption of larger gas clouds, are entrained into the outflow by hot wind. The entrainment and subsequent mixing with the hot wind both accelerates these initially non-filamentary clouds through momentum transfer and stretch them into filamentary structures (\citealt{2018MNRAS.480L.111G, 2022ApJ...924...82F}). 

Our picture aligns better with previous simulations of giant molecular cloud disruption at resolutions of 0.01-0.1 pc (see, e.g., \citealt{Kim2018, li2019disruption, 2021MNRAS.506.5512F}). These studies demonstrate that feedback from massive stars ionizes and expels low-density gas within GMCs, while intermediate and high-density filamentary structures can survive. Though our simulation's resolution is coarser than these studies, limiting the ability to resolve finer structures on parsec scales, similar behavior can be captured in our simulation: the pre-existing intermediate and high-density filamentary structures in the ISM can survive the early stage of SNe feedback. At later times, they are entrained into the outflow and gradually accelerated through momentum transfer during mixing with the hot wind, all while largely retaining their filamentary morphology (\citealt{2018MNRAS.480L.111G,2020MNRAS.498.2415M,2022ApJ...933..120B,2022ApJ...924...82F}).

The multiphase wind comprises not only hot and cool phases but also a considerable amount of warm gas. Some of the warm gas likely originates from the disk, while the rest may form within the mixing layers between the hot and cool phases. Additionally, stellar feedback, primarily from core-collapse SNe, effectively enriches the ISM in the central region. By $t=15$ Myr, a substantial fraction of the multiphase gas in the sub-kpc wind has been enriched to metallicity exceeding solar. Subsequently, even more gas in the wind attains super-solar metallicity.

After the superbubble foam breaks out, the second stage, the formation of a kpc-scale multiphase wind, takes another 10-15 Myr. From t = 15 Myr onward, the wind continues to expand outward. This multiphase outflow consists of volume-filling hot gas, mass-dominant cold and cool gas, and warm gas. As shown in Figure \ref{fig:launch_all}, the multiphase outflow extends beyond 1.5 kpc by t = 20 Myr. By t = 25 Myr, the outflow in the fiducial run FD reaches the upper boundary of the simulation volume (2 kpc along the minor axis), with numerous cool filaments stretching beyond 1 kpc. The outflow base has a radius of approximately 1.5 kpc. While some gas may escape the simulation volume by t = 30 Myr, the limited volume size and simulation time restrict our ability to fully capture the long-term evolution of the outflow.

To further explore the origin and evolution of the cool and warm phase in the outflow, particularly the formation and evolution of filamentary structures between $t=20$ to $30 \rm Myr$, when these phases in the outflow develops most rapidly, we identify filaments using a method based on the eigenvalues of the Hessian matrix of the gas density field. This approach is commonly applied to filaments identification in both the cosmic web and the ISM \citep{2007MNRAS.375..489H, Bond2010, Schisano2014, 2017ApJ...838...21Z}. In our analysis, grid cells with at least two positive eigenvalues are classified as belonging to filamentary (or core) structures and are further separated into two temperature components: neutral filaments ($T < 8000\rm K$) and $\rm H\alpha$-emitting filaments ($8000\rm K < T < 23000\rm K$). Figures~\ref{fig:fila_neutral} and \ref{fig:fila_ha} present the edge-on density field of the identified neutral filaments and $\rm H\alpha$-emitting filaments of the FD simulation from $t=20$ to $30\rm Myr$, both figures are averaged over $200\,\rm pc$-thick slice. 
Numerous filaments can be observed in both figures, while only a few dense cores appear. 

Figures~\ref{fig:fila_neutral} and \ref{fig:fila_ha} show that most cool filaments originate in intermediate to high density regions of the gas disk (below $z < 200$ pc), as highlighted in the zoom-in view of Figure~\ref{fig:fila_neutral}. Numerous filaments, ranging from tens to several hundred parsecs in length, emerge from the disk, extend into the outflow region ($z > 200$ pc), and assemble into web-like structures. Once launched into the wind, these filaments are partially accelerated and stretched by the surrounding high-velocity hot and warm gas. In this process, portions of the filaments can be reshaped into arc-like morphologies, reminiscent of the structures observed by JWST \citep{Lopez2025}. Figure~\ref{fig:overlap} further illustrates the spatial correlations between neutral filaments, $\rm H\alpha$ filaments, and bubbles of high-velocity hot gas. Many of the $\rm H\alpha$ arcs are connected to neutral filaments, yet they often reside at the shock fronts of hot winds, suggesting that at least some $\rm H\alpha$ arcs originate from interactions between cool filaments and hot outflows.

\begin{figure*}
    \centering
    \includegraphics[width=1.7  \columnwidth]{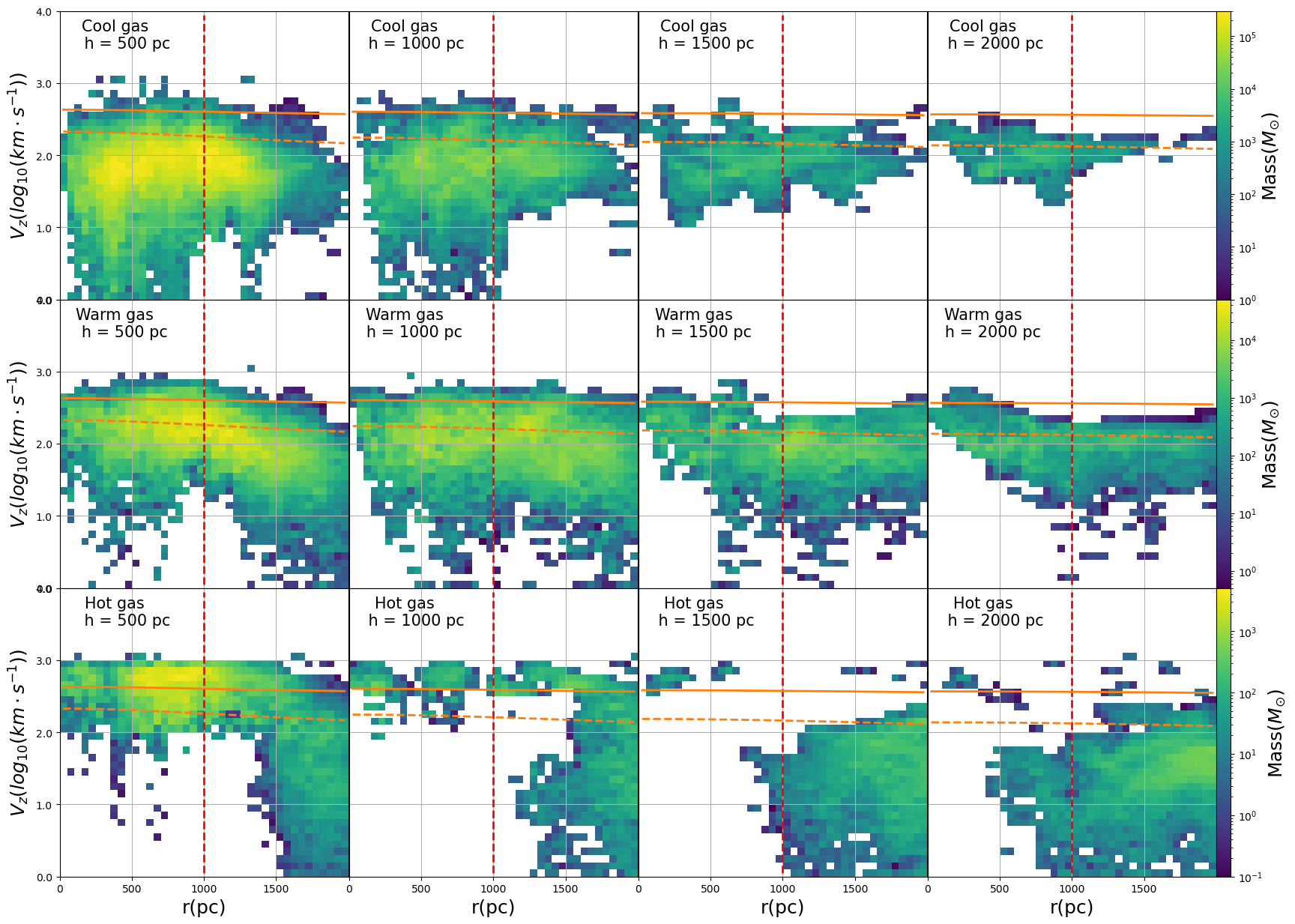}    
    \caption{The outward velocity distribution of cool ($\rm{T}<2\times 10^4\,\rm{K}$, top), warm ($2\times 10^4\,\rm{K}<\rm{T}<2\times 10^5\,\rm{K}$ middle) and hot ($\rm{T}>2\times 10^5\,\rm{K}$, bottom) gas at $t = 30 Myr$ at different distance, r, from the minor axis. From the left to the right columns, results at a height of 500, 1000, 1500, and 2000 pc are shown. The color indicates the mass outflow rate in each r-$v_z$ bin. The solid and dashed orange lines denote the escape velocity calculated with and without DM halo, respectively. The red vertical line indicates the half mass radius of the gas disk.}
    \label{fig:vz_phase}
\end{figure*}

\begin{figure*}
    \centering
    \includegraphics[width=1.7  \columnwidth]{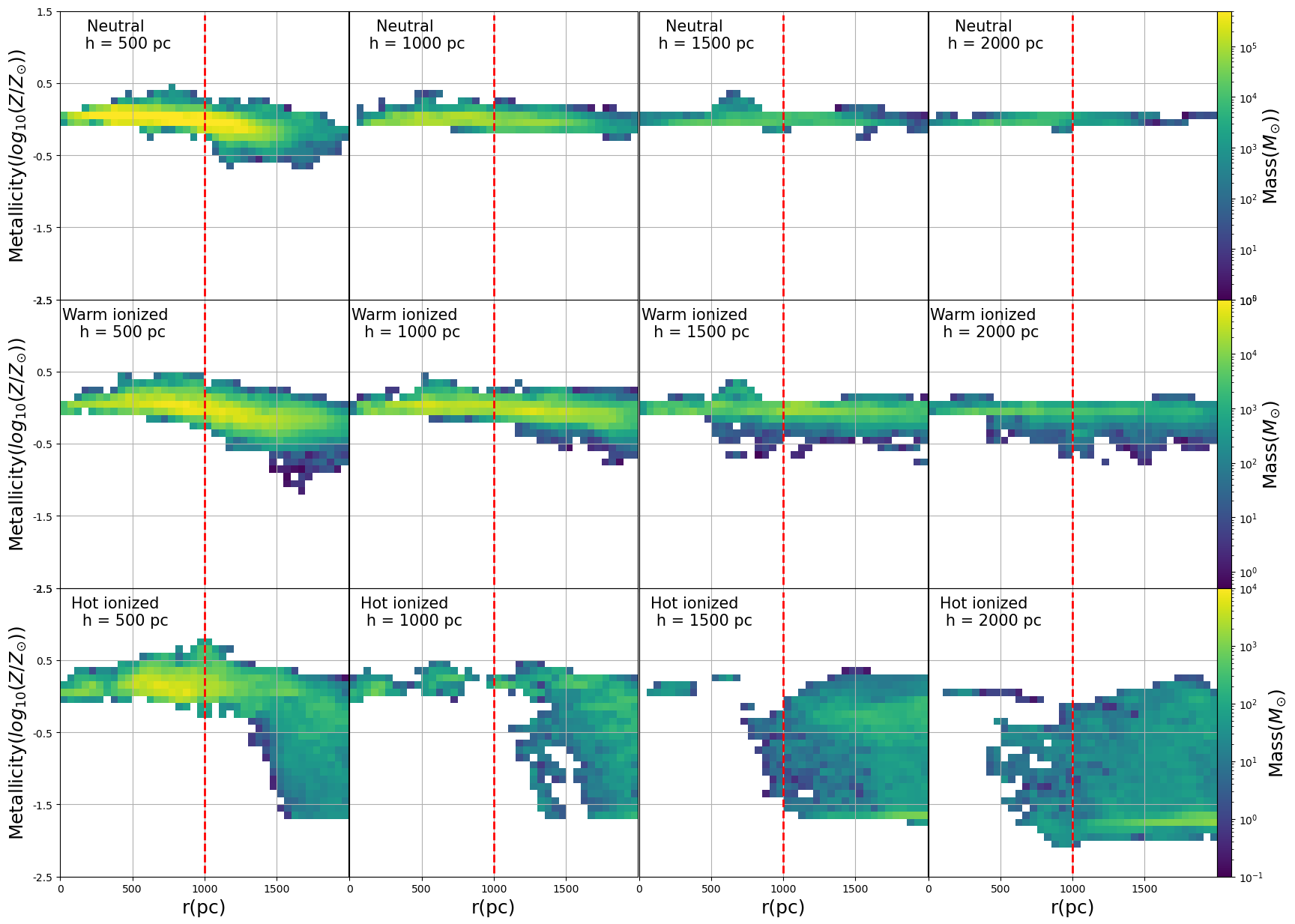}
    \caption{Similar to Figure \ref{fig:vz_phase}, shown is the metallicity distribution of the three phases at $t = 30 Myr$ at different position in the outflow region.}
    \label{fig:metal_phase}
\end{figure*}

\begin{figure*}
    \centering
        \includegraphics[width=0.48 \columnwidth]{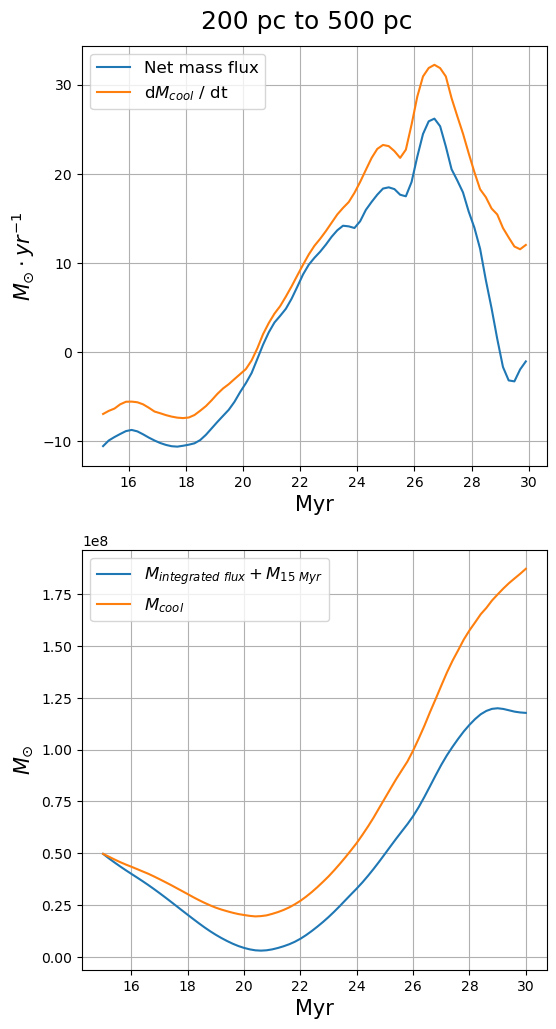}
        \includegraphics[width=0.47 \columnwidth]{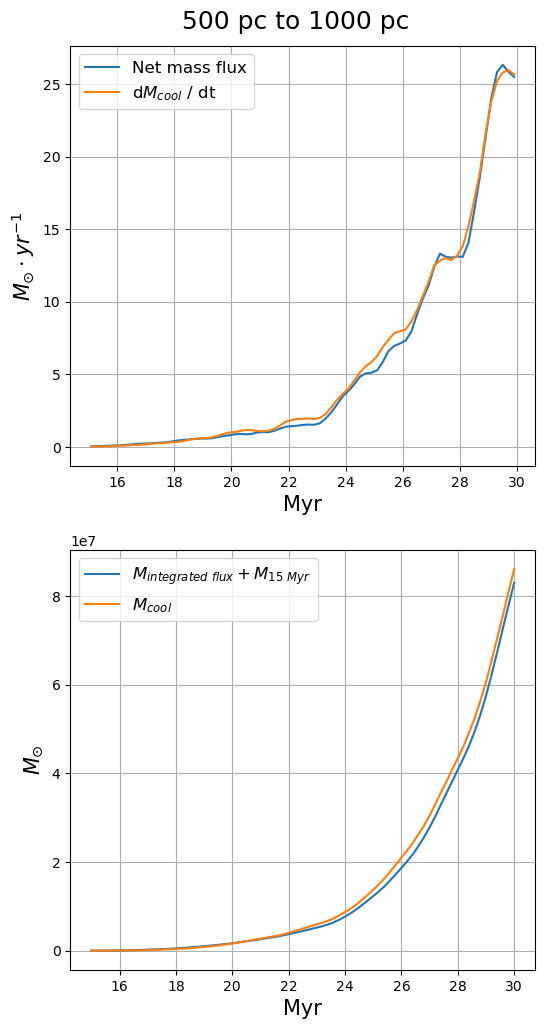}
        \includegraphics[width=0.47 \columnwidth]{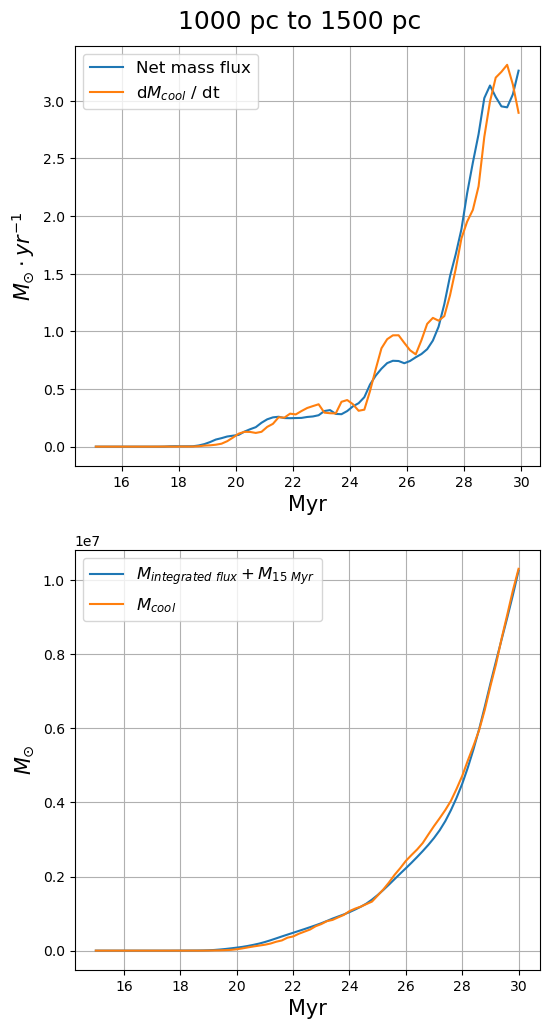}
        \includegraphics[width=0.48 \columnwidth]{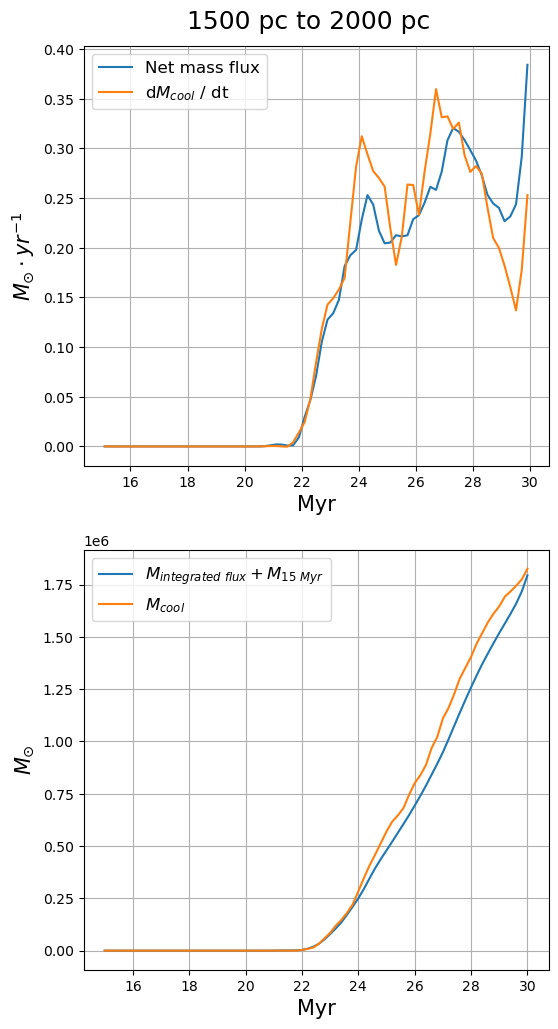}
    \caption{Top: Blue lines show the net mass flux of cool gas ($T < 2 \times 10^4\,\rm K$) into four vertical zones (200–500 pc, 500–1000 pc, 1000–1500 pc, and 1500–2000 pc) as functions of time, while the yellow lines show the  corresponding rate of change of cool gas mass in each zone. Bottom: same as top, but showing the cumulative net flux of cool gas since $t=15\,\rm Myr$ plus the cool gas mass at $t=15\,\rm Myr$ (blue), compared with the total cool gas mass (yellow).} 
    \label{fig:neu_growth}
\end{figure*}

\subsection{Evolution of different phases in the wind}
The origin of the cool and cold gas in galactic winds remains a topic of debate. Some studies propose that the cold phase originates from the cooling of the hot ionized phase (e.g. \citealt{Thompson2015}), while others suggest that it arises directly from accelerated fragments ejected from the disk (e.g. \citealt{Fujita2008, Paul2018}). These cold phases are then further accelerated through interactions with the hot phase. Our simulations demonstrate that most of the cool and cold gas in the wind originates directly from the disk. In the following subsections, we delve deeper into the evolution of different gas phases in the wind.

Figure \ref{fig:vz_phase} shows the distribution of the velocity along the minor axis, denoted as $V_z$, for the cool ($\rm{T}<2\times 10^4\,\rm{K}$, top panel), warm ($2\times 10^4$ to $2\times 10^5$ K, middle panel), and hot ($\rm{T}>2\times 10^5$ K, bottom panel) gas phases at different distances from the minor axis, r, in the FD simulation at 30 Myr. The solid and dashed orange lines denote the escape velocity calculated with and without DM halo, respectively. From left to right, the panels show results at heights of 500, 1000, 1500, and 2000 pc. A significant difference in the $V_z$-r space from z=500 to z=2000 pc is observed between the cool gas and the other two phases. At $z=500$ pc, most of the cool gas has a vertical velocity between $\sim 60\, \rm{km \cdot s^{-1}}$ and $\sim\, 300 \rm{km \cdot s^{-1}}$, which increases moderately with height. This velocity is well below the escape velocity, suggesting that most of the cool gas in the wind will eventually fall back onto the disk.

In comparison, \cite{Paul2018} found that the line-of-sight velocity of most HI in the M82 wind is around $150 \, \rm{km\, s^{-1}}$ at $\rm{z}=0.5$ kpc and $50 \, \rm{km\, s^{-1}}$ at $\rm{z} = 1\, \rm{kpc} $. Considering that M82 has an inclination angle of $77 \pm 3 ^{\circ}$, these correspond to vertical velocities $V_z$ around $550\, \rm{km\, s^{-1}}$ and $200 \sim 300 \, \rm{km\, s^{-1}}$, respectively. Observations of H$\alpha$ emission indicate that the three-dimensional radial outflow velocity of gas with a temperature of $10^4$ K is roughly 150 km/s at $r\sim 0.5$ kpc, rising to 500-600 km/s at $r\sim 1.0$ kpc, and remaining relatively constant in the outer regions (\citealt{xu2023radial}). Our simulated velocities for the cool phase are approximately $30\%-50\%$ of the observed values. 

In contrast, the warm and hot gas in our simulations exhibits higher vertical velocities. Most warm gas has velocities between $200 \sim 600\, \rm{km \cdot s^{-1}}$, while hot gas velocities range from $400 \sim 1000\, \rm{km \cdot s^{-1}}$. A significant fraction of the warm and hot ionized gas in the wind can escape the galaxy's potential well. However, the velocities of the hot gas in our simulations are still about $20\%-30\%$ lower than those inferred from X-ray observations of M82 (\citealt{2009ApJ...697.2030S}).

We also analyzed the metallicity distribution of the three phases in the outflow. As shown by Figure \ref{fig:metal_phase}, the three phases in the outflow exhibit similar metallicity distributions, with most gas having a metallicity of $Z \sim 0.5-2.0\, Z_{\odot}$. The hot gas at $z=500$ pc has a moderately higher metallcity. This level is approximately 50-100 times higher than the initial metallicity in the FD simulation. These features indicate that all gas in the outflow has undergone significant metal enrichment. As the second columns in Figure \ref{fig:fd_zoomin_face} and \ref{fig:fd_zoomin} show that, the ejecta of SNe have effectively enhanced the metallicity of cool, warm and hot phases in the central region of disc by $t=15.0$ Myr, when a significant portion of gas in disc began to flow outward to the wind.

To further confirm the origin of the cool gas in the outflow, we calculated the net mass flux of cool gas entering four height zones along the minor axis, 200-500 pc, 500-1000 pc, 1000-1500 pc and 1500-2000 pc, as a function of time. Figure \ref{fig:neu_growth} shows that the net mass flux of cool gas into the 500-1000 pc and 1000-1500 zones closely matches the growth rate of cool gas within them between $t=17.5$ and $t=30$ Myr. However, the change in cool gas mass slightly deviates from the net mass flux in the 1500-2000 pc zone, which could because of cooling of warm and hot gas. In the interface zone between disc and outflow ($200\, \rm{pc}\leq\,z<500\, \rm{pc}$ ), the mass change rate of cool gas also closely tracks the net mass flux between $t=20$ and $t=25$ Myr, but is higher at other times. 

The lower panels of Figure \ref{fig:neu_growth} show that, since $t = 15$ Myr (when the outflow begins to develop), the change in total cool gas mass between $z = 500$ pc and $z = 1500$ pc closely tracks the integrated net mass flux of cool gas. In contrast, the change in cool gas mass is slightly larger than the integrated net flux in the uppermost zone ($z = 1500$–2000 pc) and moderately larger in the lower zone ($z = 200$–500 pc). This indicates that cooling from the warm and hot phases contributes little between $z = 500$ and $z = 1500$ pc, likely offset by ionization heating of the cool gas. By comparison, the contributions from in-situ cooling, which here refers to the condensation and precipitation of warm gas during its transport and mixing \citep{Thompson2015}, in the interface region ($z = 200$–500 pc) and in the uppermost zone are at the level of around $10$ to $30\%$ and around $10\%$, respectively. 

Note that the cooling time of warm phase, fo gas with a number density of order $1\ cm^{-3} $ is around several to tens of thousands years, much shorter than the typical timescale of outflow. Therefore, both condensation and precipitation could occur efficiently in the warm phase. Conversely, processes such as stripping, heating, and ionization within the mixing layer can heat some of the cool gas to warm and hot phases. Together, these mechanisms can drive substantial mass exchange between the cool and warm phases; however, if the two processes approximately balance, the net change in the mass of each phase may remain small (\citealt{2020ApJ...895...43S, Abruzzo2022}). In the present work, we did not employ passive scalars to track disk-origin cool gas, and therefore cannot directly quantify the instantaneous transfer rate between the cool and warm phases caused by heating, mixing, or cooling. Overall, our results suggest that the total mass of the cool phase in the outflow increases only modestly during transport. Any net transfer of material from hotter to cooler phases appears to be minor and may be largely offset by heating and ionization.

\subsection{Overall feature of the outflow in various simulations}
\begin{figure*}
    \centering
    \includegraphics[width=1.7  \columnwidth]{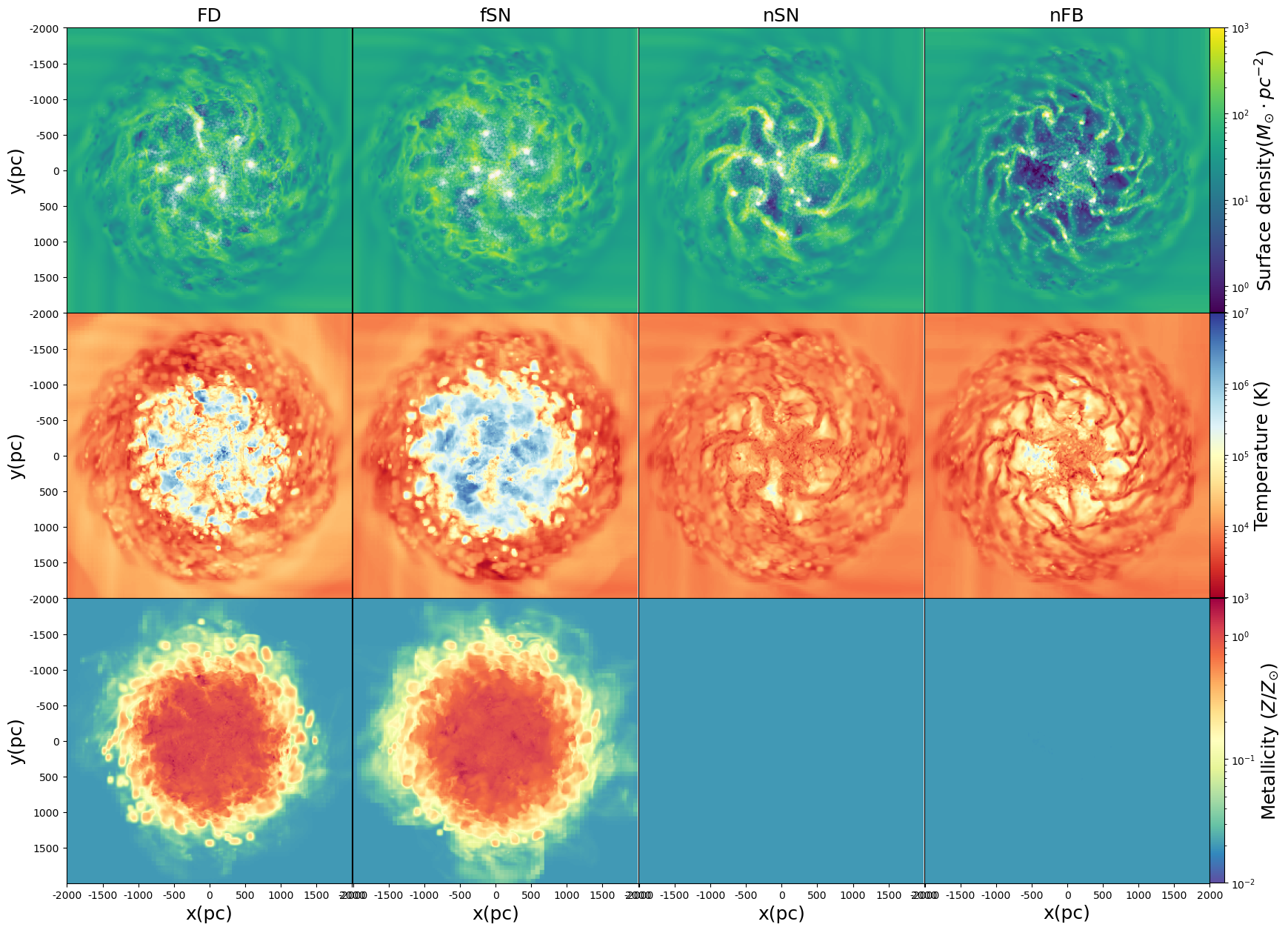}
    \caption{Face-on view of different simulations. From the top to the bottom row, column density images of the four major simulations taken at $t = 25 \rm{Myr}$, the corresponding density-weighted average temperature, and density-weighted average metallicity.}
    \label{fig:wind_comp_face_on}
\end{figure*}

\begin{figure*}
    \centering
    \includegraphics[width=1.7  \columnwidth]{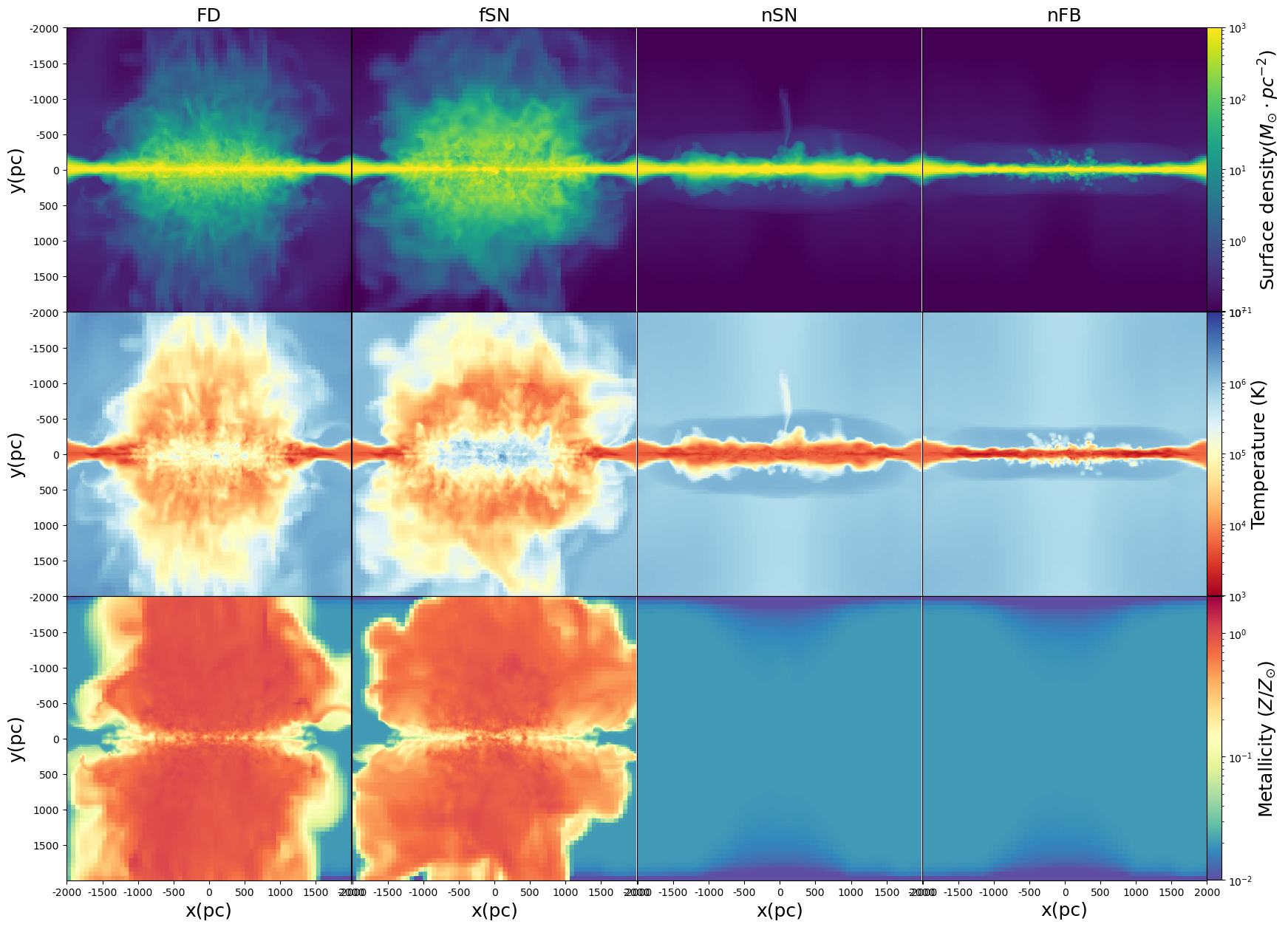}
    \caption{Edge-on view of different simulation, similar to Fig \ref{fig:wind_comp_face_on}}
    \label{fig:wind_comp_edge_on}
\end{figure*}

The launch of galactic winds varies significantly across different simulations. To investigate the impact of different feedback mechanisms, Figures \ref{fig:wind_comp_face_on} and \ref{fig:wind_comp_edge_on} compare the face-on and edge-on views of the wind and disk at t = 25 Myr for simulations with different feedback configurations.

Compared to the FD simulation, the wind in the fSN simulation occupies a slightly larger volume and has a higher gas density, due to the greater total SN energy in fSN, which ejects more gas from the disc into the outflow. However, the intense SN feedback in the fSN simulation results in a more dispersed gas distribution in the central 1 kpc region. On the contrary, the nSN simulation fails to launch a galaxy-scale outflow. In addition, the gas disk morphology in the nSN simulation is dominated by prominent cool filamentary structures, resembling the fragmented state driven by gravitational instability described by \cite{Johnson2003}. In contrast, disks in simulations with SN feedback are more turbulent, exhibiting well-developed density fluctuations across scales from tens of parsecs up to kiloparsecs. Although some smaller bubbles filled with warm gas, driven by stellar winds and radiation feedback, are present in the disc of nSN simulation, only a small amount of gas can escape into the halo, reaching heights of a few hundred to 1 kpc (Figure \ref{fig:wind_comp_edge_on}).

\begin{figure*}
    \centering
    \includegraphics[width=1.65\columnwidth]{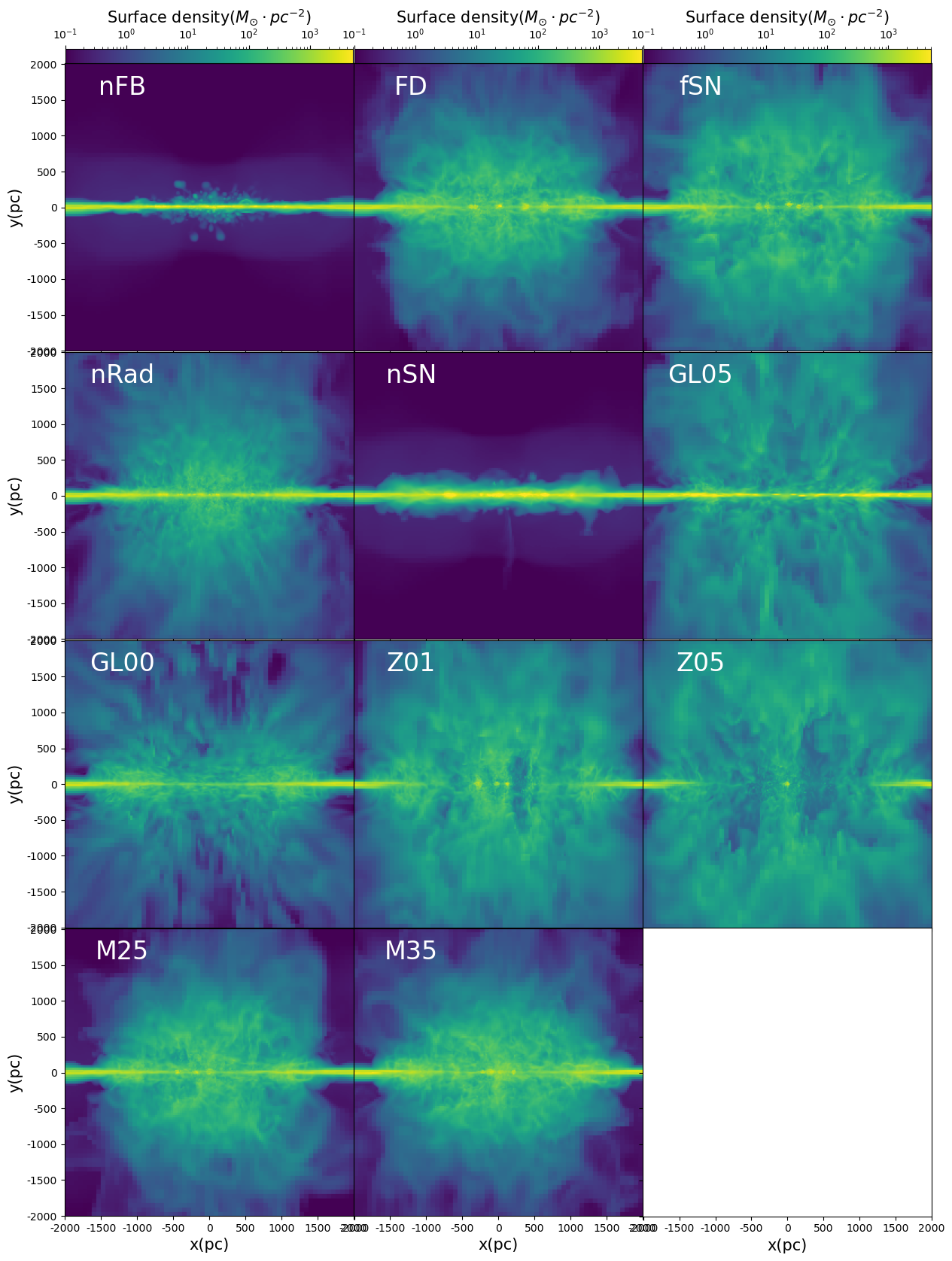}
    \caption{Edge-on view of the projected gas density at 30 Myr of all the simulations presented in the main text.}
    \label{fig:wind_edge_all}
\end{figure*}

When both stellar wind and radiation feedback are further excluded (nFB simulation), the gas disk becomes even more compact. In this case, no gas from the disk can escape into the halo. Additionally, without SNe, the gas retains its initial metallicity. Figure \ref{fig:wind_edge_all} presents edge-on views of the outflow at 30 Myr for all simulations. Reducing the gas return fraction has a noticeable effect on the outflow morphology. When the fraction is decreased to 0.5, the outflow region becomes more extended, and further reduction to 0 leads to the appearance of low-density voids within the outflow. A higher initial gas metallicity results in a larger outflow volume due to enhanced star formation and SN feedback. Meanwhile, a less massive gas disk has a moderate impact on the morphology of the outflow. 

\begin{figure*}
    \centering
    \includegraphics[width=1.8  \columnwidth]{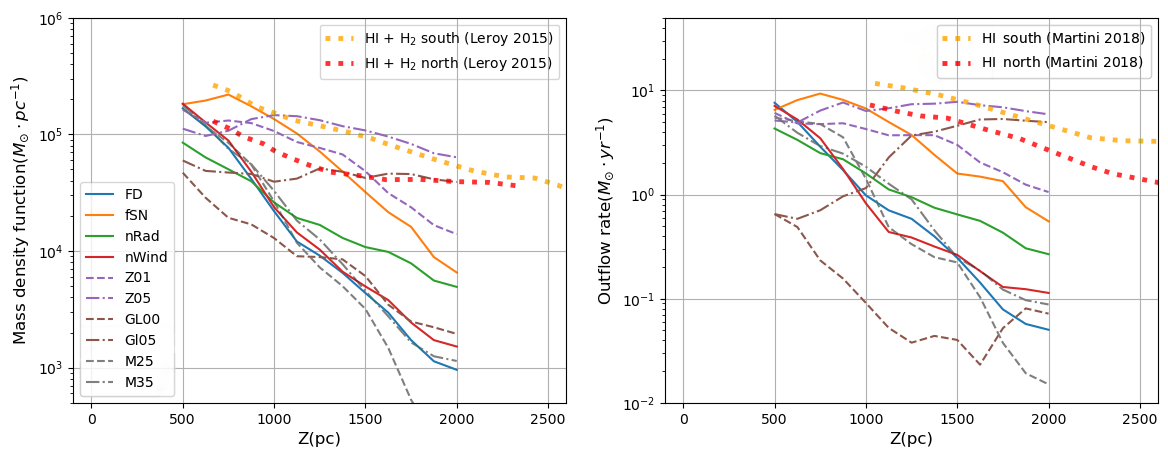}
    \caption{Left: the mass density function at $t=30$ Myr calculated along the minor axis in various simulations. Right: outflow rate at different heights along the minor axis. Each data point represents an average over a slit spanning the full box length and width, with a height of $\Delta z = 150,\rm pc$.}
    \label{fig:outf2height}
\end{figure*}

\subsection{Mass and energy outflow rate}
The mass and energy outflow rates are crucial properties of galactic winds, as they quantify a galaxy's ability to redistribute baryons and exhibit distinct features in various feedback models. Although many previous simulations have successfully launched galactic outflows in M82-like galaxies, they often struggle to reproduce the observed mass and energy outflow rates of M82. Even recent studies by \cite{2020ApJ...895...43S} and \cite{2024ApJ...966...37S} moderately underestimate the mass outflow rates of the cool and hot phases compared to observational estimates of the wind in M82 (e.g., \citealt{Adam2015, Paul2018}).

We first examine the mass outflow rate of cool ($T<2\times 10^4\, \rm{K}$) gas, covering the temperature range of molecular, atomic, and ionized hydrogen. \cite{Adam2015} derived the mass density function of both HI and $\mathrm{H}{2}$, defined as mass per unit length, by integrating the mass within $\pm 1.5\, \rm{kpc}$ along the minor axis. For comparison, we calculate the mass density function of cool gas at $t=30$ Myr in our simulations within a 3.0 kpc-wide slit along the minor axis, shown in the left panel of Figure \ref{fig:outf2height}. Note that HI and $\mathrm{H}{2}$ gas have temperatures below 8000 K; thus, the density of the cool ($T<2\times 10^4\, \rm{K}$) phase in M82 should be moderately higher than the values in \cite{Adam2015}, since ionized hydrogen typically has lower density than neutral hydrogen. Furthermore, we measure the mass outflow rate within a narrower 1 kpc-wide slit and compare it with the results of \cite{Paul2018}, which consider only HI and should be regarded as a lower limit for the cool gas outflow rate, in the right panel of Figure \ref{fig:outf2height}. Because the temperature ranges in our simulations differ from those in \cite{Adam2015} and \cite{Paul2018}, readers should interpret these comparisons with caution.

The HI+$\mathrm{H}_{2}$ mass distribution and mass outflow rate at $\rm{z}\gtrsim 1000$ pc in simulations using the variable SNe model are approximately 1.5 dex lower than observational estimates. In contrast, simulations with higher energy injection rates, whether from increased initial metallicity or the fixed SNe energy model, can drive cool gas more efficiently, better matching the observed outflow rates below z = 1500 pc. Notably, a higher initial metallicity of $0.5 Z_{\odot}$ yields a cool gas outflow rate comparable to observations throughout the simulation volume. Meanwhile, a less massive gas disk has a mild impact on the outflow rate of cool phase.

On the other hand, Section 3 indicates that simulations with a fixed SNe energy model or higher initial gas metallicity produce a starburst higher than estimation of M82's recent starburst by $\sim 30\%-100\%$. To simultaneously replicate both the starburst and outflow characteristics, we need to improve the efficiency of coupling SN energy to the ISM. One potential solution is to incorporate the enhanced feedback effects of clustered supernovae (e.g. \citealt{Gentry17,2018MNRAS.481.3325F}). In our subsequent work (\citealt{2025ApJ...982...28L}), we explored the impact of clustered SNe on outflow properties. The inclusion of enhanced feedback from clustered SNe successfully reproduced both the total stellar mass of the starburst and the outflow rates observed in M82.

In addition to the amount of SNe energy injected, the gas supply in the nucleus region of the disk also plays an important role in shaping the properties of outflow. Disabling gas return from star particles drastically reduces the mass of cool gas below 1 kpc and decreases the cool gas outflow rate by a factor of 10 throughout the simulation volume. Meanwhile, a $50 \%$ gas return fraction yields a cool gas outflow rate comparable to observations above 1.5 kpc, but significantly lower (by a factor of 10) at 0.5 kpc. One potential explanation is that a substantial portion of gas is accreted by sink/star particles, depleting the gas supply in the starburst core. Initially, this reduces resistance to the outflow, but over time it limits the availability of cool gas to fuel the wind and suppresses star formation after $t=10$ Myr (see Figure \ref{fig:sfh}), thereby reducing the total number of SNe injected into the ISM. Consequently, disabling gas return can enhance the outflow rate in the early stage while suppressing it at later times. In addition, the evolution of outflow rates is not synchronized in different simulations, which can be shown by the results below. 

\begin{figure*}
    \centering
    \includegraphics[width=1.8  \columnwidth]{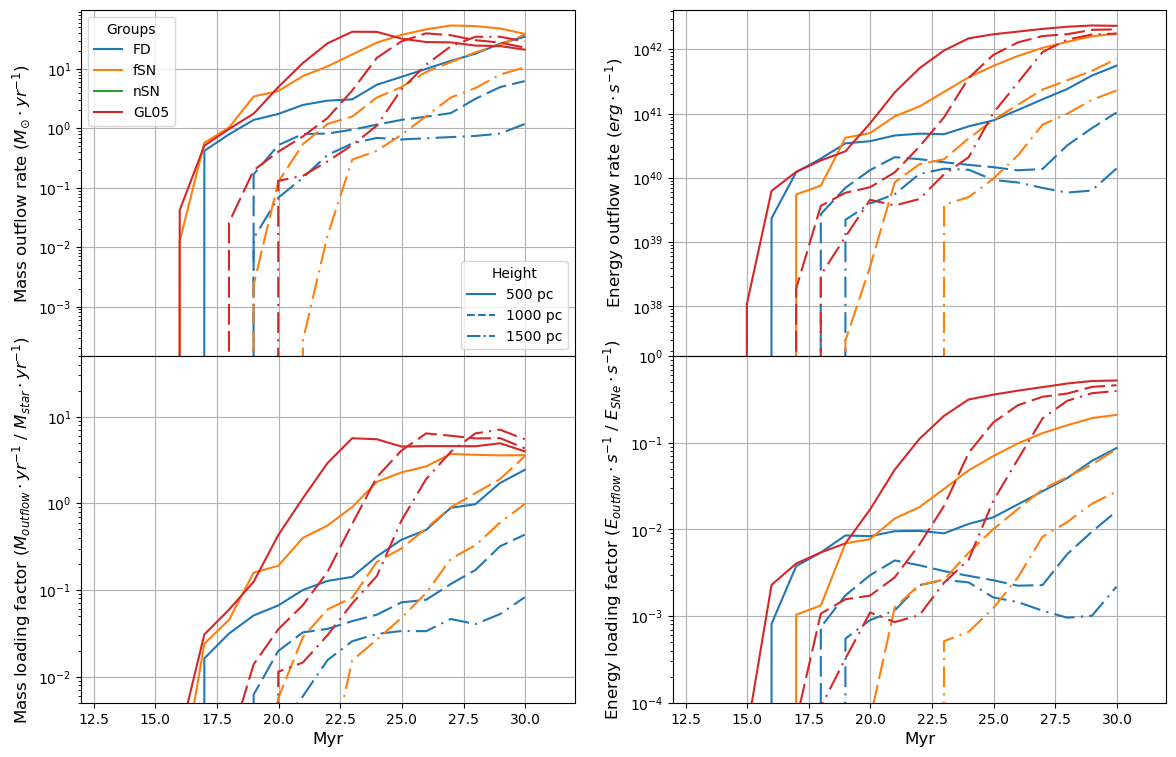}
    \caption{Left: mass outflow rate (top) and mass loading factor (bottom) from various simulations, measured at the height of 500, 1000, and 1500 pc as a function of time. Right: energy outflow rate (top) and energy loading factor (bottom). No detectable galactic wind is produced in nSN, with all corresponding values remaining below the lower limit. Note that, owing to the rapid decline in the star formation rate, the calculated mass loading factor  at later times may be overestimated.}
    \label{fig:mass_load}
\end{figure*}

We calculated the total mass and energy outflow rates for all three gas phases as functions of time in each simulation, along with the mass and energy loading factors, which are crucial properties of galactic outflows. The mass and energy loading factors, defined as:
\begin{equation}
    \eta_{M} = \frac{\dot M_{out}}{\rm{SFR}}
\end{equation}
and
\begin{equation}
    \eta_{E} = \frac{\dot E_{out}}{\dot E_{\rm{SNe}}}
\end{equation}
where $\dot M_{out}$ and $\dot E_{out}$ are the total mass outflow rate and energy outflow rate respectively. Here, both $\dot M_{out}$ and $\dot E_{out}$ are evaluated for all thermal phases across the entire radial extent along the major axis, not just the confined region shown in Figure \ref{fig:outf2height}. The results are presented in Figure \ref{fig:mass_load}. In simulations with SN feedback, the mass loading factors at z = 0.5 kpc and z = 1.0 kpc are approximately 2-5 and 0.8-5, respectively, at the end of the simulation. The current star formation rate in M82 is uncertain, ranging from 2-10 $\rm{M_{\odot} yr^{-1}}$ (e.g. \citealt{1998ApJ...498..541K,Adam2015}). The total mass outflow rate in M82 is estimated to be $\sim 20\,\rm{M_{\odot} yr^{-1}}$ at heights of $\rm{z}\sim 1-2$ kpc, including the cold, cool, warm, and hot phases (\citealt{2009ApJ...697.2030S,Adam2015,Paul2018}). This implies a mass loading factor of 1-10 at z = 1 kpc, which is broadly consistent with our simulation results. Note that the nSN simulation has never produced a detectable outflow.

The mass loading factor for starburst galaxies with given properties, such as stellar mass, exhibits significant variations across different simulations. This discrepancy can be attributed, in part, to inconsistencies in the methods used to calculate outflow rates. For instance, FIRE-2 simulations employ a spherical shell approach with radii of $0.10\ r_{vir}$ and $1.0\ r_{vir}$ (\citealt{2021MNRAS.508.2979P}), while the study based on the EAGLE simulation utilize a more complex Lagrangian particle tracking method (\citealt{2020MNRAS.494.3971M}). In contrast, the LYRA project adopts a conical approach with a height equal to the simulation box size and a width defined by the star-forming region (\citealt{Thales2021}). These different methodologies can lead to substantial variations in the calculated mass loading factors. We also notice that many simulations have a much longer simulation time than in our work, allowing galaxies to evolve to a stable state. Therefore, a longer time interval is used in the calculation of the mass and energy loading factor, which can reduce the fluctuations in these measured values. 

On the other hand, uncertainties exist in observational estimates of the SFR and mass outflow rate. For example \cite{2023MNRAS.518.4084Y} suggest that the outflow rates of cold and cool gas in M82 could be uncertain by several times. Furthermore, uncertainties in M82's inclination angle can introduce systematic errors. To improve the predictive power of simulations, further observational and theoretical investigations are necessary to reduce these uncertainties and refine our understanding of individual galaxies.

\section{Discussions}
\subsection{Comparison with previous simulation}

The multiphase outflows driven by starbursts in M82-like galaxies have been extensively investigated over the past two decades (e.g., \citealt{2008ApJ...674..157C, 2013MNRAS.430.3235M, 2020ApJ...895...43S, 2024ApJ...966...37S}). The principal improvement of this study is the self-consistent treatment of star formation and feedback. In our framework, stars form within dense, gravitationally collapsing clumps that naturally emerge from the evolution of the galactic disc. Their formation and subsequent feedback are modeled with the sink particle method, directly linking star formation to the underlying evolution of gas disc. In contrast, earlier studies generally prescribed the starburst properties, such as intensity, temporal evolution, spatial distribution, and SN feedback, through imposed assumptions rather than outcome of disc evolution.

For example, \cite{2008ApJ...674..157C} adopted fixed injection rates of $10^{42}$ $\rm erg/s$ and $1$ $\rm M_{\odot}/yr$, distributing energy and mass according to the initial gas density to approximate star clusters. \cite{2013MNRAS.430.3235M} also used a fixed energy injection rate of $10^{42}$ $\rm erg/s$, but varied the number of SSCs and the duration of star formation. In contrast, \cite{2020ApJ...895...43S} employed a variable SN injection rate tied to an idealized starburst history: stars formed at $20$ $\rm M_{\odot}/yr$ from 5–35 Myr and $5$ $\rm M_{\odot}/yr$ from 35–70 Myr, with clusters of $10^{7}$ $\rm M_{\odot}$ confined within $r<1$ kpc. Their model produced a total feedback energy of $\sim6\times10^{57}$ erg, corresponding to an average of $\sim6\times10^{42}$ $\rm erg/s$ during 5–35 Myr. More recently, \cite{2024ApJ...966...37S} revised the SFH to a distributed burst model, allowing for broader cluster distributions and variable masses, while maintaining a comparable average injection rate of $\sim6\times10^{42}$ $\rm erg/s$ over 40 Myr.

By $t=30$ Myr, when outflows are well developed, our simulations form $3.5\times10^8$–$6.0\times10^8$ $\rm M_{\odot}$ of stars, with the lower end consistent with observational estimates of $2.0\times10^8$ to $3.5\times10^8$ $\rm M_{\odot}$ \citep{1993ApJ...412...99R, Schreiber2003}. The associated SN energy input is $2.5\times10^{57}$ to $4.3\times10^{57}$ erg, yielding an average injection rate of $2.6\times10^{42}$ to $4.5\times10^{42}$ erg/s. This is a few times higher than the rates in \cite{2008ApJ...674..157C} and \cite{2013MNRAS.430.3235M}, but only $40\%$ to $75\%$ of those adopted by \cite{2020ApJ...895...43S} and \cite{2024ApJ...966...37S}.

A key advantage of our approach is that stars form naturally within dense gas clouds, rather than being imposed externally. This allows us to resolve the early stages of outflow launching in greater detail, capturing how stellar wind and SN bubbles expand, interact with, and eventually break out of their dense birth environments. Our simulations also reveal how isolated bubbles merge to drive the large-scale flow and how SNe ejecta mix with and enrich the surrounding multiphase ISM. In addition, by incorporating gas return from GMCs, we provide a more self-consistent description of ISM evolution.

Moreover, the development of cool filaments in our simulations shows some notable features that have not been revealed by previous simulations of the galactic winds. In a number of previous studies (e.g. \citealt{2008ApJ...674..157C}), cold filaments in the outflow were attributed to interactions between the hot phase and small, dense clouds, which are fragments of giant molecular clouds in the central disk, that were entrained into the wind. These clouds were initially non-filamentary, but subsequent mixing with the hot flow both accelerated them through momentum transfer and stretched them into filamentary structures \citep{2018MNRAS.480L.111G,2022ApJ...924...82F}.

In our simulations, consistent with many previous studies, most of the cool gas in the outflow originates from the disk. However, unlike several earlier works, we find that most of the cool gas in filaments was initially part of filamentary ISM structures in the disk. These pre-existing filaments survive the early stages of supernova feedback, are subsequently entrained into the outflow, and are gradually accelerated through momentum transfer during mixing with the hot wind, all while largely retaining their filamentary morphology.

As shown in Figures \ref{fig:fd_zoomin_face} and \ref{fig:fd_zoomin}, numerous cool filaments and several dense cores within the disk remain less affected by SN feedback than the diffuse ISM, enabling them to survive for 10–15 Myr rather than being rapidly heated or ionized. During this period, SN-driven bubbles in the starburst region expand and merge to form a superbubble, after which filamentary structures are expelled from the disk at $t \sim 15$ Myr, following the outflow of the hot and warm phases. Figures \ref{fig:fila_neutral} and \ref{fig:fila_ha} further illustrate how these filaments escape the disk, interact with other phases in the interface region ($z=200$–500 pc), and evolve into web-like networks and $\rm H\alpha$-emitting arcs, features reminiscent of those observed by \cite{Lopez2025}. Our results highlight survival and transport as the dominant pathway.

Our picture of the survival of cool filaments in the starburst region is consistent with previous high-resolution simulations of the destruction of individual GMCs, with a resolution of about 0.01-0.1 pc. Those simulations show that a considerable fraction of the high-density filamentary structures in the GMCs can survive the feedback from newly born massive stars (see, e.g., \citealt{Kim2018, li2019disruption, 2021MNRAS.506.5512F}). 

Furthermore, our quantitative analysis (Figure \ref{fig:neu_growth}) shows the net transfer of material from hotter phases into the cool phase provides only a secondary contribution to the total cool gas mass in the outflow. This net transfer occurs primarily within the interface region between the gas disc and the outflow (200 pc <z< 500 pc) and the uppermost zone. In contrast, the dominant net source of cool gas in the outflow is the transport of pre-existing cool ISM from the central disc region. Nevertheless, mass exchange between phases is still possible: material can condense from hotter phases into the cool phase, but this gain is likely counterbalanced by heating, ionization, or mixing-driven transfer of cool gas into warmer phases. 

Nevertheless, we speculate that, in addition to filamentary structures spanning tens to 100 pc, compact, roughly spherical gas clumps, only a few to ten parsecs in size and containing cool or cold gas in the nuclear region prior to the starburst, could also survive stellar feedback and become entrained in the multiphase outflow, as shown in previous simulations. However, our simulations show almost no such small clouds remaining in the starburst region. This likely reflects two factors: (i) star particles form near the density peaks of clumps, where feedback most effectively disrupts their compact cores, and (ii) our resolution is insufficient to fully resolve clouds with only a few parsecs across. Consequently, the contribution of compact, dense clouds to cool filament formation in the wind may be underestimated in our simulations.

\subsection{Limitation of this study}
We note that there are several limitations in our current work. First, the recipes and parameters adopted in several processes, such as the star formation, various feedback mechanisms, and gas return, are not yet firmly constrained by theory or observations, and therefore require further verification. The properties of the star clusters and the effect of clustered SN would be different if these recipes and parameters were changed. Second, our model of M82 has not incorporated the tidal interactions between the three members in the M81 group. This omission may cause the initial conditions in our simulations to deviate from reality, particularly in the central region. Third, the initial mass model is derived from present-day observations and assumes a fully developed ISM at the onset of the starburst. Consequently, the adopted initial conditions may not reflect the actual pre-burst environment. We have shown that the initial mass and metallicity moderately influence both the star formation history (SFH) and outflow properties. These factors are expected to alter the timescale and peak of the starburst, as well as the quantitative properties of the outflow, but not the overall evolutionary picture. Fourth, the side length of our simulation box is 4 kpc, which is unable to track the development of wind at a larger radius, and the current results also may have suffered from some boundary effect.

Resolution is another important factor. Higher resolution would allow us to capture faster fragmentation in dense star-forming clumps, resolve pc scale dense cores and the free expansion phase of individual SNRs, and reduce artificial cooling. Our convergence tests show that lower resolution delays the rise of the SFR, slightly prolongs the starburst, and affects the formation of cold and warm structures such as filaments and arcs. As demonstrated by \citet{Cooper2009}, the number of fragments formed from cloud break-up increases with resolution. Thus, at higher resolution, the properties of filamentary and arc-like features in the outflow may differ from our current results to some extent. A higher-resolution study is therefore needed to confirm our conclusions. Lastly, our current simulation only treats the stars formed in previous starbursts as a gravitational potential background, while neglecting the feedback from those stars. With a total mass around $3-5 \times 10^{9} M_{\odot}$, the older stars could potentially prevent clustering and suppress star formation to a certain degree through type Ia supernovae and IR radiation.

Furthermore, there are several shortcomings in the results of our simulations, with respect to the observations of the recent nuclear starburst and outflow in M82. The total stellar mass formed in the starburst in many of our simulations is higher than the value estimated from observations. However, this issue can be resolved if the average gas return fraction of star-forming clumps is reduced to $0\%-50\%$ or with a reduced initial gas mass of $2.5 \times 10^{9} \rm{M_{\odot}}$. Moreover, the outflow velocity in our simulations is slower than in the observations. This situation is worse for the cool gas, which is only one-third to half of the observed value. In addition, the outflow rate of the cool gas in many simulations is lower than that observed at the height $z\gtrsim 1$ kpc. Although our simulations can form super star clusters, the limited resolution is likely unable to capture the enhanced feedback efficiency due to clustered SNe as demonstrated in previous studies (\citealt{Gentry17,2018MNRAS.481.3325F}), probably due to overcooling. 

In the second paper of this series, we carry out simulations in a large volume and with a different recipe to couple the SN feedback energy to the neighbouring ISM, and to return the gas in sink/particles to the ISM. We find that a better agreement with the observations on the total stellar mass and outflow rate can be obtained by moderate revisions on the related modules and the inclusion of the enhanced feedback effect because of clustered SNe. More details can be found in \cite{2025ApJ...982...28L}. However, we find that further improvement is needed to increase the outflow velocity of cool and cold gas in the wind. 

\section{Conclusions}

In this work, we conducted high-resolution hydrodynamical simulations to investigate the recent starburst and the subsequent launch of a galactic-scale wind in M82. Unlike previous studies that relied on prescribed starburst models, we self-consistently resolved the star formation process using a sink particle module. We explored the impact of various factors, including different feedback mechanisms, gas return from sink/star particles into the ISM (destruction of star-forming gas clouds), initial gas metallicity, and gas disk mass, on the starburst and outflow. Our findings are summarized below.

\begin{enumerate}
\item Our simulations successfully reproduce a starburst in the nuclear region, lasting 20-25 Myr with a peak SFR of $20-45\, \rm{M_{\odot}\, yr^{-1}}$. The total stellar mass formed in the starburst mostly ranges from $5\times 10^8 \,\rm{M_{\odot}}$ to $6\times 10^8\, \rm{M_{\odot}}$, which is however moderately higher than the value in M82 inferred from observation. However, reducing the gas return fraction to less than $50\%$ or adopting a lower initial gas disk mass of $2.5\times 10^9\, \rm{M_{\odot}}$, can bring the total stellar mass broadly in line with observations. The observed relation between surface star formation rate and gas surface density in our simulations is several times higher than the classical Kennicutt-Schmidt law but is comparable to that of highly active starburst galaxies.

\item The super star clusters formed in our simulations exhibit a cumulative cluster mass function that agrees reasonably well with observations at the high-mass end. However, the model underpredicts the number of lower-mass clusters. The median integrated star formation efficiency of star particles ranges from $10\%$ to $30\%$, which agrees with previous simulations but exceeds observational estimates of $2\%$ to $10\%$. The metallicity of stars formed in the burst varies from 0.02 to several times solar metallicity, with a mean value of 0.1-0.2 solar for an initial gas metallicity of 0.02 solar. Increasing the initial metallicity to 0.1 solar yields a mean stellar metallicity closer to observed values in M82's nuclear region

\item The launch of the multiphase galactic-scale wind in our simulations occurs in two stages. The first stage, lasting about 10 Myr, involves the formation and breakout of a superbubble foam composed of hot, warm, and cool filamentary gas. Initially, combined radiation, stellar wind, and SN feedback create numerous small bubbles. The feedback ionizes and heats low-density ISM, while leaving many intermediate to high density filaments relatively intact. As the bubbles expand and merge, they form a large superbubble that contains multiphase gas, including embedded cool filaments, which eventually breaks out of the disk. The second stage, spanning 10-15 Myr, involves the development of the kpc-scale multiphase outflow. Hot gas expands outward at the highest velocity, followed by warm gas, and then cool gas in filaments. Interactions between the hot phase and cool filaments stretch the filaments. Transport of pre-existing cool ISM from the starburst region is the primary net contributor to the total mass of the cool phase in the outflow. In contrast, mass transfer from hotter phases, such as condensation in the boundary layer or precipitation driven by radiative cooling, appears to be largely offset by the concurrent transfer of material from the cool phase to warmer phases during mixing. As a result, these processes make only a minor net contribution to the total cool mass in the outflow, acting primarily within the interface region between disc and outflow ($z=200$–500 pc) and the uppermost vertical zones ($z>1500$ pc).  

\item The outflow velocity of cool gas in our simulations is approximately $30-50\%$ of the observed values in M82. Similarly, the velocities of warm and hot gas are about $20-30\%$ lower than observed. When the galactic wind is fully developed, the mass outflow rate of cool gas in simulations with a total injected SN energy of $\sim 4-6 \times 10^{57} \rm{erg}$ is comparable to the observations within the range $500<z<2000$ pc. However, simulations with lower SN energy input result in significantly lower cool gas outflow rates. The mass loading factor in most simulations ranges from 0.8 to 5, consistent with observational estimates for M82. To better match both the outflow velocities and rates observed in M82, enhanced feedback efficiency due to clustered SNe is necessary, while simultaneously reducing the total stellar mass formed in the simulation.

\item SN feedback acts as the primary driver of the outflow, while gas return significantly influences the starburst and outflow properties. The initial mass of gas disc has a moderate effect, while stellar wind and radiation feedback have minor effect on the starburst and the properties of outflow. Radiation feedback can suppress the formation of super star clusters that massive than $10^7 \,\rm{M_{\odot}}$.

\end{enumerate}

\section*{Acknowledgements}
We are grateful to the anonymous reviewer for his/her insightful comments and suggestions. We thank the helpful discussions with Antonios, Katsianis and Cheng, Li. This work is supported by the National SKA Program of China (Grant Nos. 2022SKA0110200 and 2022SKA0110202), by the National Natural Science Foundation of China (NFSC) through grants 11733010 and 12173102, and by the China Manned Space Program through its Space Application System. The authors would like to thank National Supercomputer Center in Guangzhou
for providing high performance computational resources. Post simulation analysis was performed on the HPC facility of the School of Physics and Astronomy, Sun Yat-Sen University.

\section*{Data Availability}
The data underlying this article will be shared on reasonable request to the corresponding author.



\bibliographystyle{mnras}
\bibliography{Main}




\appendix

\section{details of mass model}
\subsection{Stellar component}

Observations in the 1970s and 1980s had identified M82 as a dusty disk galaxy with a central core. Later observations further revealed the existence of more complex structures.  For instance, \cite{Telesco1991} identified characteristics of a bar in the central region of M82. The same structure was confirmed by \cite{Wills2000}. In addition, some studies revealed the existence of spiral arms (\citealt{Mayya2005}) in M82. However, in this work, we omit the effect of bar and spiral arms, only account for a stellar fraction from a bulge, or in some contexts stellar spheroid (\citealt{Tomisaka1992, Strickland2000}), and an exponential disk. The scale height and scale length of both the bulge and stellar disk have been constrained by several observations (\citealt{Ichikawa1995, Forster2001};\citealt{2006ApJ...649..172M}). 

The work of \cite{Levy2024} and \cite{Lim2013} highlight a significant difference in the spatial distribution of stars formed in different starburst events in M82. The younger stars formed in the most recent starbursts we aim to simulate primarily concentrated in the nuclear region. In contrast, stars formed during earlier starbursts are dispersed more widely across the whole disk region, some of which reach as far as a few kpc. Based on the two-part SFH for stars located in the inner 3 kpc disk derived by \cite{Mayya2005}, which consists of a continuous and moderate star formation lasted for over 10 Gyr with an SFR under $0.3\, \rm{M_{\odot} \cdot yr^{-1}}$ and a brief starburst began around 0.8 Gyr ago and lasted for 0.3 Gyr with a peak SFR of $20\, \rm{M_{\odot} \cdot yr^{-1}}$, we estimate that around $5.9 \times 10^{9}\, \rm{M_{\odot}}$ of stars were formed during this period, of which approximately $3.3 \times 10^{9}\, \rm{M_{\odot}}$ of low to intermediate-mass stars remained till the beginning of the starburst we aim to simulate.

To describe the mass distribution and the potential of the stellar disk component, we use the widely adopted \cite{Miyamoto1974} disk profile. The density profile of a Miyamoto-Nagai disk is described by
\begin{equation}
    \rho(R,z) = \frac{Mb^{2}[aR^{2}+(a+3\sqrt{z^{2}+b^{2}})(a+\sqrt{z^{2}+b^{2}})^{2}]}{4\pi[R^{2}+(a+\sqrt{z^{2}+b^{2}})^{2}]^{5/2}(z^{2}+b^{2})^{3/2}}
\end{equation}
where M is the disk mass, a is the radial scale length, and b is the vertical scale height. The corresponding gravity potential is given by
    \begin{equation}
        \Phi(R,z) = \frac{-GM}{\sqrt{R^{2}+(a+\sqrt{z^{2}+b^{2}})^{2}}}
    \end{equation}
the mass of the stellar disk is set to the $3.3 \times 10^{9} M_{\odot}$ based on the above discussion, and the scale length and scale height of the stellar disc used in our simulation are $1200$ pc and $200$ pc respectively.

As for the bulge, the scale radius and mass of stellar bulge are $r_{0} = 450$ pc and $\rm{M_{bulge}} = 6.0\times10^8 \rm{M_{\odot}}$ respectively according to \cite{Ichikawa1995} and \cite{Forster2001}. Previous studies often use the King profile to describe the mass distribution and potential of a bulge.
\begin{equation}
    \rho(r) = \frac{\rho_{0}}{1+(r/r_{0})^3/2}, 
\end{equation}
where \(r_0\) is the scale radius of the stellar bulge. The potential of which is then
\begin{equation}
    \Phi(r) = \frac{-\rm{GM_{bulge}}}{r_{0}} \left[ \frac{ln\{ (r/r_{0})+\sqrt{1+(r/r_{0})^{2}} \}}{r/r_{0}} \right]. 
\end{equation}
The bulge overlaps spatially with the recent starburst region. Therefore, a significant portion of the bulge's stellar mass likely originated from the recent starburst events we aim to simulate. Hence, unlike previous studies, we take a different approach in our simulation by incorporating the bulge as a part of the initial gas disk. By directly adding the mass of the bulge to the initial mass of the gas disk, we can obtain a gas disk profile that more accurately reflects the initial condition before the onset of the recent starburst. This method also allows for a more self-consistent formation of stars in the nuclear region during our starburst simulation. The detailed setup of the gas disk is described in the following section.
 
Note that there are some uncertainties associated with the chosen parameters for both the bulge and stellar disk. These uncertainties arise from the notable extinction and assumptions on the stellar population, initial mass function, star formation history, etc. A more accurate understanding of the stellar component may be of considerable interest, but is beyond our ability. 

\subsection{Gas disk}
Based on various observations, previous studies suggest that molecular gas is the dominant gas component in M82, followed by atomic hydrogen, and then warm and hot ionized gas. Currently, a considerable fraction of the molecular and atomic gas lies beyond the gaseous disc, due to the interaction with M81 and the outflow driven by the wind. Since our aim is to replicate the processes of starburst and outflow, and the interaction is relatively difficult to reproduce, we assume that most of the molecular and atomic gas belongs to the gaseous disk at the beginning of the simulation, i.e. about 20-50 Myr before the present epoch. 

The mass of $\rm{H}_2$ in M82 is usually inferred from the observations of the CO line, where the CO-$\rm{H}_2$ conversion factor, $X_{\rm{CO}}$, has a noticeable uncertainty. The total mass of $\rm{H}_2$ in M82 at the present time varies from $1.3\times 10^9$ to $2.4\times 10^9$ $\rm{M_{\odot}} $ in different studies, depending on the adopted value of $X_{\rm{CO}}$ (\citealt{2002ApJ...580L..21W, 2013PASJ...65...66S, Adam2015, 2021ApJ...915L...3K}). Most of the $\rm{H}_2$ resides on the disk and streams. The molecular gas mass in the outflow comprises about $25\%-40\%$ of the total molecular gas in M82. 
Meanwhile, the total mass of $\rm{H}_2$ mass in the nucleus region is around $2.0 \times 10^8 \rm{M_{\odot}}$ (\citealt{Wild1992}; \citealt{Naylor2010}). In addition, 
\cite{Adam2015} found a scale length of 0.6 kpc along the major axis and 0.5 kpc along the minor axis for the molecular gas in the disk of M82.

Observation conducted by \citet{Richard1978} suggested a total HI mass of around $5.0 \times 10^8 \rm{M_{\odot}}$ in M82. A similar result was obtained by \citet{Appleton1981}, suggesting a total HI mass of $7.2\times 10^8 \rm{M_{\odot}}$. More recent observations indicate that the HI mass in M82 is between $7.5$ and $8.0 \times 10^8 \rm{M_{\odot}}$ ( \citealt{Yun1999, 2008AJ....135.1983C, DeBlok2018}). \cite{Adam2015} provides a detailed mapping of the HI distribution and found a scale length along the major axis to be 2.6 kpc.

In this work, we do not explicitly solve the chemistry of the gas component. Namely, we do not resolve the atomic, molecular, and ionized gas separately. Accounting for the uncertainties in the estimated mass of molecular gas and some of the gas in the nuclear region that had turned to stars in the past 20-50 Myr, we adopt a gas disc with total mass $\rm{M_{gd}} \approx 2.5, 3.0, 3.5 \times 10^9 \rm{M_\odot}$ respectively in different runs in our simulation.  

Most of the previous simulations of galactic wind in M82-like galaxies uses a single \cite{Miyamoto1974} gas disk (e.g., \citealt{Strickland2000, 2008ApJ...674..157C}). More recently, \cite{Paul2018} proposed a different setup of the mass model to obtain the analytical form of the gravity potential. In their work, a triple Miyamoto-Nagai disk model (\citealt{Smith2015}) was employed. Though the \cite{Smith2015} model may produce an accurate fit to the mass distribution of an exponential gas disk, it fails to reproduce the gas profile of M82. \cite{Adam2015} found the density profile of HI in M82 best fitted by a projected \(\rho \propto r^{-2}\) density profile, and $\rm{H_{2}}$ best fitted by a projected $\rho \propto r^{-3.5}$ density profile along the major axis. We employ a similar method to that described by \cite{Smith2015}, using three Miyamoto-Nagai disks to fit the projected density profile found by \cite{Adam2015}. Our choice of parameters for the triple Miyamoto-Nagai disk model is given in Table \ref{tab:3mdisk}. 

\begin{table}
	\centering
	\caption{The parameters for the triple Miyamoto-Nagai disk model used in this work, where $h$ is the scale height of the disk. Combining the three disk components results in a thick disk that corresponds to a radial density profile \( \rho \propto r^{-2} \) with ellipticity e=0.3.}
	\begin{tabular}{lccr}
		\hline
		Disk component & $b/h$ & $a/h$ & $m/m_{tot}$\\
		\hline
		1 & 1.0 & 2.39 & 1.37\\
		2 & 1.0  & 2.67 & 0.0779\\
		3 & 1.0  & 4.83 & -0.449\\
		\hline
	\end{tabular}
	\label{tab:3mdisk} 
\end{table}

\subsection{Dark matter halo}
Early research of the M82's rotational curve indicated that its rotational curve profile was likely to be Keplerian (\cite{Gotz1990}). Consequently, many previous simulations of galactic wind in M82-like galaxies did not include the gravitational contribution from the dark matter halo. However, more recent observations of the CO line and star clusters suggest a different scenario. \cite{2012ApJ...757...24G} found the rotational curve of M82 is considerably flattened towards the outer region, which is commonly interpreted as evidence for the existence of the dark matter halo (DM halo hereafter). However, the complexity of interactions and mergers among the three members of the M81 group complicates efforts to accurately constrain the parameters of the DM halo. In \cite{Oehm2017}, three-body simulation and Markov chain Monte Carlo (MCMC) method are employed to constrain the initial parameters of the three DM halos, and a high-resolution N-body simulation was conducted to further investigate the evolution of the triplet. They demonstrated that the current DM halos, in which M82 and its neighboring galaxies are embedded, arise from the distortion, blending, and merging of the original DM halos. Simple analytical models centered at M82 may struggle to reproduce the true profile of the galaxy's DM halo. For the sake of simplicity, we adopt a modified NFW profile in this work. The density distribution of the original NFW dark matter halo is given by 
\begin{equation}
    \rho(r) = \frac{\rho_{0}}{\frac{r}{R_{s}}\left( 1 + \frac{r}{R_{s}}\right)^{2}}, 
\end{equation}
and the potential is given by
\begin{equation}
    \Phi(r) = \frac{-4\pi G\rho_{0} R_{s}^{3}}{r} ln \left( 1 + \frac{r}{R_{s}} \right).
\end{equation}
where \(R_{s}\) is the scale radius and \(\rho_{0}\) is the characteristic density, obeying the following relationships
\begin{equation}
	R_{s} = R_{vir}/c,  
\end{equation}
where c is the concentration parameter.

Though the NFW profile has been widely adopted, it is not without its shortcomings, particularly in the scenario we are considering. The density of an NFW DM halo does not converge to a finite value towards the center, while observations suggest that many galaxies have a constant-density core in the central area of their DM halos. Such a discrepancy is known as the cusp-core problem. \cite{Read2016} in their simulation shows that star formation and stellar feedback activities occurring in the embedded dwarf galaxy can transform the initial NFW DM halo into a 'core-NFW' profile with a constant density core in the starburst region ($<R_{1/2}$) within a Hubble time. The profile of a 'Core-NFW' halo can be described by 
\begin{equation}
    M_{c\rm{NFW}}(<r) = M_{\rm{NFW}}(<r) \cdot f^{n},
\end{equation}
in which $f = tanh(\frac{r}{r_{c}})$. In the case of M82, the galaxy has experienced multiple starburst events and developed a significant galactic scale outflow; a fully developed core-NFW profile is the adequate choice, which requires $n = 1$. The radius of the core $r_{c}$ takes the radius of the inner starburst region, i.e., $r_{c} = 500 pc$. To further simplify the mass model and thus obtain an analytical expression of the potential, we used an NFW profile patch with a constant density core in the \(r<R_{1/2}\) region.

We have also adopted the $c-\rm{M_{vir}}$ relation proposed by \cite{Maccio2007} to further reduce the number of undetermined parameters of the DM halo, which states
\begin{equation}
    \rm{log}\ c = 1.071\pm 0.027 -(0.098 \pm 0.009) ( \rm{log M_{vir} - 12}).
\end{equation}  
With the $c-\rm{M_{vir}}$ relation enforced, two free parameters remain: $\rm{M_{vir}}$ and $R_{vir}$. We used the Markov Chain Monte Carlo(MCMC) method to find the optimal combination of the two parameters that best fit the observed rotation curve of \cite{2012ApJ...757...24G}. To obtain a more accurate result, we excluded the central 200 pc region, where the velocity dispersion is significantly higher than the local rotational velocity. Additionally, we also excluded the velocity bump located $\sim 500$ pc at the eastern part of M82, which is caused by an SSC designated as cluster 'z' by \cite{McCrady2007}. We find the best fitting halo mass to be $\rm{M_{vir}}=6.0\times 10^{10} \, \rm{M_{\odot}}$, and the best fitting virial radius is $R_{vir} = 53\, \rm{kpc}$.  Fig \ref{fig:rotcurve} shows that the overall rotation curve of our model closely matches the observed result of \cite{2012ApJ...757...24G}.

\section{Refinement scheme}
\label{sec:smr}

\begin{figure}
    \centering
    \includegraphics[width=1.0\linewidth]{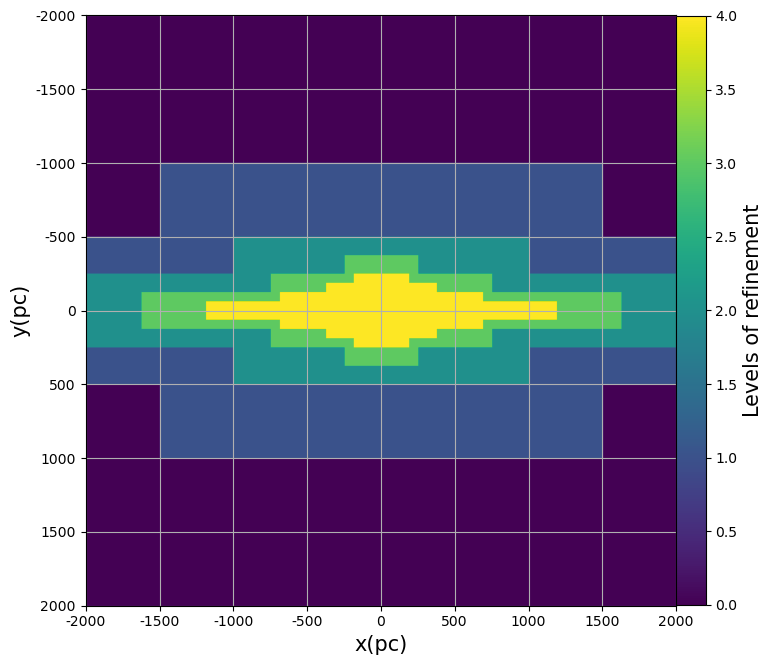}
    \caption{The refinement map of our simulations. The resolution is given by $\rm{\Delta x} = 64/2^{refinement\ level}$.}
    \label{fig:refinement}
\end{figure}

In our simulations, we adopted a static mesh refinement scheme. Resolving the evolution of the gas disk, the starburst, and supernova injection ideally requires parsec-scale resolution and even better. Constrained by computational resources, we use a maximum resolution of $\Delta x = 4$ pc in the central, high-density region of the gas disk. The resolution then decreases gradually with distance from the center, reaching $64$ pc in the outer regions of the domain. There are four levels of refinement. The refinement map is shown in Figure \ref{fig:refinement}.

\section{Convergence test}

\begin{figure}
    \centering
    \includegraphics[width=1.0  \columnwidth]{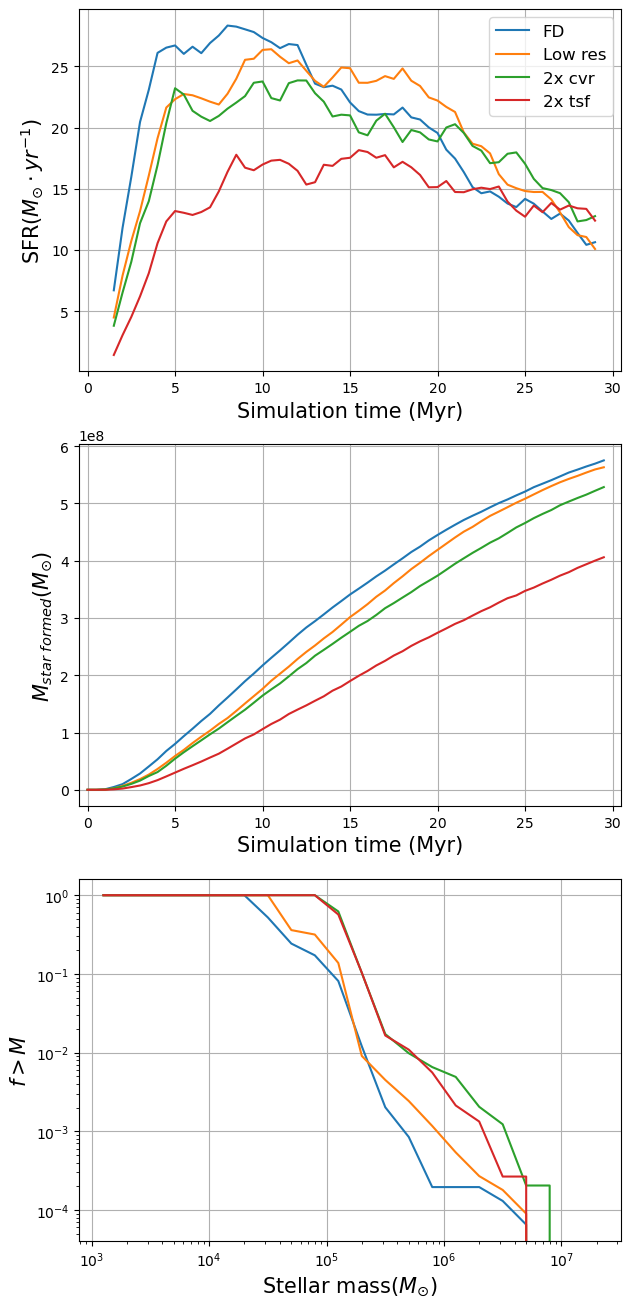}
    \caption{From the top to the bottom: star formation history, the integrated SFH, and the cumulative cluster mass function in simulation simulation FD, Low res, $\rm{2\times}$ cvr, and $\rm{2\times}$ tsf (see Table \ref{tab:setup_convergence} for more details).} 
    \label{fig:sfh_cnv}
\end{figure}

\begin{figure*}
    \centering
    \includegraphics[width=1.8  \columnwidth]{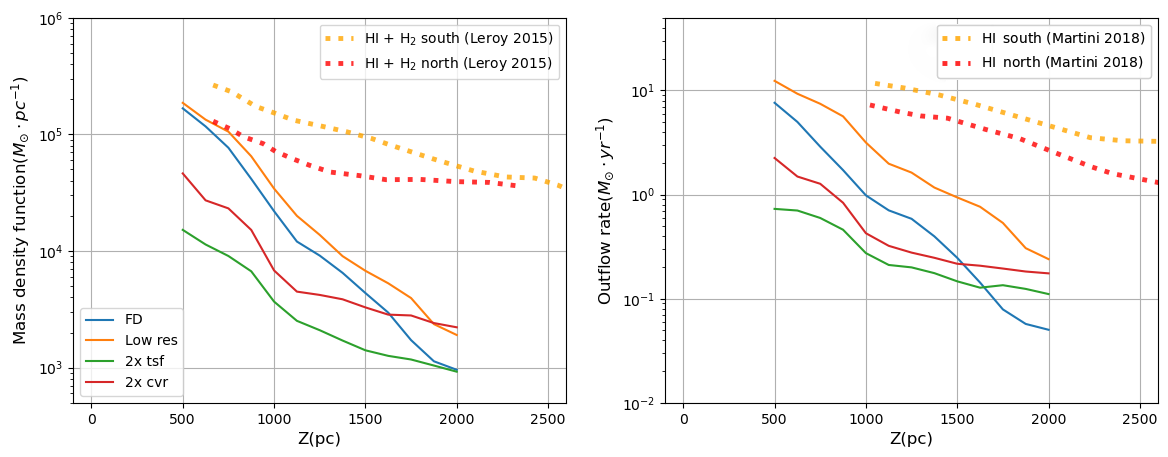}
    \caption{Same as Fig \ref{fig:outf2height}, the mass density function (left) and outflow rate at different heights (right) of the four simulations.}
\end{figure*}

To comprehensively assess our results, it is essential to examine the convergence of star formation-related recipes and modules across different resolutions. A key variable in the sink particle-based star formation recipe is the control volume radius, which influences both accretion and star formation rates. As resolution changes, the physical size of the control volume also changes. Additionally, the volume affected by feedback energy depends on the grid size. Therefore, different spatial resolutions can affect star formation and the impact of stellar feedback. Coarse simulations may not capture the free expansion phase of SN remnants, leading to the blending of SN ejecta with the ISM and lowering the core temperature to the thermally unstable regime ($10^{4} \sim 10^{6} K$). This can result in significant energy dissipation before expansion, artificially suppressing the effects of SNe.

To assess the impact of resolution and other parameters on our results, we conducted additional simulations with the setups listed in Table \ref{tab:setup_convergence}.

Figure \ref{fig:sfh_cnv} presents the SFH and stellar mass distribution for different resolutions and parameter variations. Lower resolution simulations exhibit a slower initial rise in SFR, reaching the peak SFR approximately 2 Myr later than the fiducial run. Despite this difference, the final total stellar mass remains similar to the fiducial run. Expanding the control volume slightly suppresses the SFR by $10-20\%$ and reduces the total stellar mass by $10\%$. Meanwhile, increasing the star formation timescale significantly reduces the peak SFR by $40 \%$ and the total stellar mass by $30 \%$. Additionally, the SFH profile for the longer star formation timescale shows a slower decline after the peak, suggesting potential for further star formation.

\begin{table}
	\centering
	\caption{Simulations run for convergence test. $R_{cv}$ indicates the radius of control volume for each sink particle, in unit of the grid size dx. }
	\begin{tabular}{lccr}
		\hline
		Simulation name & central resolution & $R_{cv}$ & $t_{\rm{sf}}$\\
		\hline
		Low res & 8 pc & 1.5 dx & $t_{ff}/0.2$\\
		2x cvr & 8 pc & 3.0 dx & $t_{ff}/0.2$\\
        2x tsf & 8 pc & 1.5 dx & $t_{ff}/0.1$\\
		\hline
	\end{tabular}
	\label{tab:setup_convergence} 
\end{table}

Altering the spatial resolution and control volume size has a modest impact on the SFH, resulting in approximately $10-20\%$ variations. The overall profile of the SFH and the total stellar mass remain largely similar to the fiducial run, indicating that the SFH is relatively insensitive to these parameters. In contrast, adjusting the star formation timescale significantly reduces the SFR, leading to a flatter SFH profile. Therefore, accurately representing the star formation process within molecular clouds necessitates a careful choice of the star formation timescale.

The development of the outflow appears to be more sensitive to changes in spatial resolution and control volume size than the SFH. The HI outflow rate in the low-resolution run exceeds that of the fiducial run by a factor of 1.5 at z = 500 pc and by a factor of 4 at the edge of the simulation domain. Several factors may contribute to this discrepancy. First, lower resolution leads to SNe energy being injected into a larger volume, potentially resulting in greater radiative cooling losses during the free expansion phase but also more efficient acceleration of surrounding cool gas through direct energy injection. Second, lower resolution may limit the formation of unstable structures, preventing hot, ionized gas from escaping through gaps and channels between clumps, thus enhancing the efficiency of SN-driven outflows. Third, as shown in Section 4, the evolution of outflow rates is not synchronized in different simulations.

Expanding the control volume significantly reduces the cool gas outflow rate at z < 1600 pc but increases it at z > 1700 pc. This behavior arises from the time-dependent nature of the outflow rate at a given height z: it initially increases and then decreases due to the properties of the SFH and associated SN feedback. However, the timing of these increase and decrease phases can vary across simulations. A larger control volume reduces the SFR, leading to a lower peak outflow rate at all heights. Nevertheless, the outflow rate in the '2x cvr' simulation can exceed that of the FD simulation in certain regions and at specific times.


\bsp	
\label{lastpage}
\end{document}